\documentclass[10pt,twocolumn,twoside,journal]{IEEEtran}

\ifCLASSINFOpdf
\else
   \usepackage[dvips]{graphicx}
\fi
\usepackage{url}

\usepackage{amsmath}
\usepackage{graphicx}
\usepackage{float}
\usepackage{multirow}
\usepackage{subfigure}
\usepackage{stfloats}
\usepackage{cases}
\usepackage{subeqnarray}
\usepackage{array}
\usepackage{enumerate}
\usepackage{amssymb}
\usepackage{url}
\usepackage{color}
\usepackage{dsfont}
 \usepackage[table]{xcolor}
\usepackage{caption}
\usepackage{soul}
\usepackage[switch]{lineno}
\usepackage{algpseudocode}
\usepackage{algorithmicx,algorithm}
\usepackage{diagbox}
\usepackage{multirow}
\usepackage{soul}
\usepackage{tabularx}

\DeclareMathAlphabet\mathbfcal{OMS}{cmsy}{b}{n}

\colorlet{mygreen}{green!60!gray}

\begin{document}
\title{Image Splicing Detection, Localization \\ and Attribution via JPEG Primary Quantization Matrix Estimation and Clustering}

\author{Yakun Niu, Benedetta Tondi, \IEEEmembership{Member, IEEE}, Yao Zhao, \IEEEmembership{Senior Member, IEEE},\\ Rongrong Ni and Mauro Barni, \IEEEmembership{Fellow, IEEE}

\thanks{Y. Niu, Y. Zhao and R. Ni are with Institute of Information Science, Beijing Jiaotong University, Beijing 100044, China, and also with Beijing Key Laboratory of Advanced Information Science and Network Technology, Beijing 100044, China (e-mail: \{niuyakun; yzhao; rrni\}@bjtu.edu.cn). (\emph{Corresponding author: Yao Zhao}.)}
\thanks{M. Barni and B. Tondi are with the Department of Information Engineering and Mathematics, University of Siena, 53100, Siena, Italy (e-mail: barni@dii.unisi.it; benedettatondi@gmail.com).}

\vspace{-1cm}
}

\maketitle
\begin{abstract}
Detection of inconsistencies of double JPEG artifacts across different image regions is often used to detect local image manipulations, like image splicing, and to localize them. In this paper, we move one step further, proposing an end-to-end system that, in addition to detecting and localizing spliced regions, can also distinguish regions coming   from different donor images. We assume that both the spliced regions and the background image have undergone a double JPEG compression, and use a local estimate of the primary quantization matrix to distinguish between spliced regions taken from different sources. To do so, we cluster the image blocks according to the estimated primary quantization matrix and refine the result by means of morphological reconstruction. The proposed method can work in a wide variety of settings including aligned and non-aligned double JPEG compression, and regardless of whether the second compression is stronger or weaker than the first one. We validated the proposed approach by means of extensive experiments showing its superior performance with respect to baseline methods working in similar conditions.
\end{abstract}

\IEEEpeerreviewmaketitle

%
%
%

\section{Introduction}

Detection of double JPEG (DJPEG) compression plays a major role in image forensics since double compression reveals important information about the past history of an image \cite{Jessica2008, Li2008}.
This is the case of image splicing detection and localization. When part of an image is spliced from a donor JPEG image into a target JPEG image to create a composite forgery (which is eventually recompressed so that the final image is double compressed), it is quite common that the compression setting used for the donor image is not equal to that used to compress the target image.
As a consequence, the original and spliced regions of the forged image exhibit different (double) compression artifacts thus providing the basis for the detection and localization of the spliced region. Most of the methods proposed so far to detect image splicing based on double compression artifacts work under the following simple assumptions:
\begin{enumerate}
\item{the tampered region (or regions) comes from a single donor image. Very few attempts have been made to identify forgeries containing multiple spliced areas coming from different donor images. Yet, given an image with several copy-pasted regions, it is possible, at least in principle, to identify the different origin of the tampered areas by recognizing that spliced regions coming from different donor images probably underwent a different double compression history;}
\item{the spliced region (or regions) is taken from a non compressed image and spliced into a JPEG image. After recompression, the forged area has undergone only a single JPEG compression (SJPEG) while the background has been compressed twice \cite{amerini2014splicing,deng2019deep,chen2011detecting}. In the following, we refer to this situation as DJPEG vs SJPEG detection. In practical scenarios, however, it is more likely that both the donor and the target images have been JPEG compressed, thus calling for the development of techniques capable to work in a DJPEG vs DJPEG setting\footnote{The DJPEG vs DJPEG scenario is often addressed indirectly assuming that the second compression of the foreground is performed on a misaligned JPEG grid, while an aligned DJPEG is applied to the background (or viceversa). In this setting, many systems implicitly regard the background as SJPEG hence reducing this case to a SJPEG vs DJPEG scenario.}. }
\end{enumerate}

In this paper, we propose a general approach to simultaneously perform image splicing detection, localization and attribution of regions coming from different donors. The proposed method, which is specifically thought to work in the DJPEG vs DJPEG scenario (but can also work in the SJPEG vs DJPEG case), relies on the estimation of the quantization matrix used in the first compression step of a DJPEG image (in the following, we will refer to such a matrix as primary quantization matrix and we will indicate it by $Q_1$). Specifically, the proposed method works by providing a local blockwise estimate of the primary quantization matrix and then clustering the image blocks according to such an estimate. Splicing detection is achieved by recognizing the presence of more than one cluster, while the exact number of clusters identifies the number of donor images used to create the forgery. Eventually, by looking at blocks belonging to different clusters, we can localize the spliced areas and attribute them to different donor images.
As an additional advantage, the proposed method also works when the second compression is stronger than the first one, e.g. when the quality factor used for the first compression ($QF_1$) is larger than that used for the second one ($QF_2$), that is when $QF_1 > QF_2$. Moreover, it can also cope with non standard quantization matrices.

\enlargethispage{\baselineskip}

For the estimation of the $Q_1$ matrix, we adopt the CNN-based estimator proposed in \cite{niu2019SPL}, due to its capability to provide a good estimation results also on  small patches. To get the tampering map with the indication of the spliced regions, we associate to each $8 \times 8$ block of the image a vector with the estimated quantization steps, then we apply Spectral Clustering (SC) to such vectors \cite{von2007tutorial}. In order to determine the number of clusters, we trained an ad-hoc  Convolutional Neural Network (CNN) taking as input the estimated quantization steps. If only one cluster is found, the image is classified as a non-forged image and no further operation is carried out. In the presence of multiple clusters, we apply SC, then the largest cluster is associated to the image background, while the others are regarded as belonging to spliced regions, each cluster corresponding to a different donor image. The tampering map and the estimated number of spliced regions are finally refined by enforcing the spatial coherence and smoothness of the clusters through morphological reconstruction.

The main contributions of our work can be stated as follows:

\begin{itemize}
  \item{We introduce a new image forensics problem, hereafter referred to as donor image attribution, or simply attribution, whose goal is to decide if two or more spliced regions within a tampered image come from the same donor image and group them accordingly.}
  \item We propose a new method to jointly localize tampered regions in JPEG images and distinguish between spliced regions coming from different donor images, under the assumption that the donor images have been compressed with different $Q_1$ matrices. We do so in the most common DJPEG vs DJPEG scenario.
  To the best of our knowledge,
   this is the first method performing localization and attribution of multiple spliced regions by relying on the analysis of compression traces. The system is based on the following ingredients: primary quantization matrix estimation, clustering and morphological reconstruction. We also designed and trained a CNN to identify the number of clusters present in the image by looking at the local estimate of the primary quantization matrix. Thanks to such a CNN, the proposed system is able to carry out splicing detection, localization and attribution simultaneously, hence providing an end-to-end system for image splicing forensics.
  \item{We carried out an extensive experimental campaign to evaluate the effectiveness of the proposed method in a wide variety of settings and tampering scenarios. In order to evaluate the performance of tampering localization in the case of spliced regions originating from multiple sources or donor images, we introduce a new metric based on the Normalized Mutual Information (NMI) metric, commonly used in pattern recognition applications to assess the performance of clustering.}
\end{itemize}

The proposed approach is a very general one since it can be seamlessly applied in a wide variety of cases. First, it is one of the few approaches explicitly thought to work in a  DJPEG vs DJPEG setting, secondly it works also when $QF_1 > QF_2$. Eventually, it maintains good performance regardless of whether the first and second compression grids are aligned or not.  Moreover, the method can be applied also when non-standard quantization matrices are used and hence the  quality factor $QF$ is not defined.

The rest of this paper is organized as follows. After a brief review of related methods (Section \ref{sec.ArtPrior}),
in Section \ref{sec.problemstatement}, we present the general tampering localization setup considered in this paper and introduce the main notations. The proposed method for image tampering localization is described in Section \ref{sec.propmethod}. Section \ref{sec.methodology} describes the methodology we followed for the experimental analysis, whose results are reported in Section \ref{sec.results}. We conclude the paper with some final remarks in Section \ref{sec.conclusion}.

\enlargethispage{\baselineskip}

\section{Prior art}
\label{sec.ArtPrior}

Several techniques for image tampering localization have been proposed in the forensic literature, relying on different manipulation traces, e.g. inconsistencies of resampling artifacts \cite{mahdian2008blind,bunk2017detection}, or presence of different sensor noise patterns \cite{chierchia2014guided,korus2017multi}.
More often, inconsistencies of JPEG compression artifacts are exploited. Due to the wide diffusion of the JPEG compression standard, in fact, image editing software often re-save the edited images in JPEG format, hence making it possible to detect and localize tampering based on the analysis of the traces of double JPEG compression and their inconsistencies across the tampered image.
In the following, we briefly review the relevant literature about DJPEG detection for image tampering localization.

It is well known that double JPEG compression leaves peculiar artifacts
in the DCT domain, in particular, in the histograms of block-DCT coefficients \cite{popescu2004statistical}. Accordingly, many tampering detection algorithms rely on the statistical analysis
of DCT coefficients \cite{Li2008,pasquini2014multiple}.
Some example of methods for detecting
double compression artifacts in the non-aligned DJPEG scenario relying on handcrafted features
computed in the pixel or the DCT domain
are described in \cite{chen2011detecting,luo2007BACM,qu2008a,bianchi2012detection}.
%
Early approaches
were designed to work on the whole image,  to detect if the analysed image has
undergone a global single or double JPEG compression. Such methods are not applicable in a tampering
detection scenario, where only part of the image has been manipulated, due to the difficulty of estimating the
required statistics on small blocks. To cope with this problem, a bunch of other methods have been developed for DJPEG localization \cite{amerini2014splicing,lin2009fast,bianchi2011improved}.
In general, these methods have low spatial resolution and
their performance drop significantly
when regions smaller than $256\times 256$ are considered.
More recently, a new class of CNN-based methods have been proposed. They are able to improve the spatial resolution of DJPEG localization and then can be conveniently used for tampering localization, (see, for instance, \cite{WZ16,barni2017aligned} - for both aligned and not-aligned DJPEG detection, and \cite{deng2019deep}, \cite{Kwon2021WACV}  - for the aligned DJPEG case).
All the above methods focus on the SJPEG vs DJPEG scenario, that is, they work under the assumption that the tampered areas are single compressed while the background is double compressed. In contrast, very few work has been done
to specifically address the more challenging DJPEG vs DJPEG scenario considered in this paper.
In principle, methods capable to estimate the primary quantization matrix of DJPEG images, e.g. \cite{Jessica2008} and \cite{Lukas2003}, could be applied to this scenario. However, the methods in \cite{Jessica2008,Lukas2003} work on the full image, and hence are not suitable for localization.
In \cite{Farid2009}, a technique is proposed to detect whether part of an image was formerly compressed with a JPEG quality lower than that used for the rest of the image  ($QF_1 < QF_2$), by means of exhaustive recompression with every quality factors.

The works that are more closely related to this paper are  \cite{bianchi2012image} and \cite{wang2014exploring}.
Both these methods estimate the $Q_1$ matrix on a local basis and output a
map with the probability that a DCT block has been double-compressed.  The method in  \cite{bianchi2012image} works under the assumption that the histograms of the unquantized DCT coefficients are locally uniform in the non tampered region. Moreover, accurate detection can be achieved when the spatial resolution is larger than
$256\times 256$ pixel. Two approaches are proposed in \cite{bianchi2012image} for the cases of aligned and non-aligned DJPEG.
%
The method in \cite{wang2014exploring} is designed for the case of aligned DJPEG compression.
%
Both methods work better when $QF_2 > QF_1$, while performance are significantly worse in the opposite case.

Being able to distinguish spliced regions coming from different donor images, our method can also be used for image phylogeny, where the identification of the donor images is a required step to identify the relationships between the spliced image and its parent images \cite{ImagePhylogeny,RochaTIFS2011,Oliveira2016ImagePhylogeny}
and use it to reconstruct the
history of semantically similar images.
From this perspective,  the goal of the system described in this paper is not very different from  that of image phylogeny applications, the main difference being that in the image phylogeny scenario the donor images are assumed to be available to the analyst.

Finally, we point that several methods based on CNNs  have been developed in the last few years addressing other tampering detection and localization problems, e.g. the problem of copy-move tampering \cite{wu2018busternet,Tin,islam2020doa}. Moreover, CNN-based approaches addressing   the problem of forgery  detection and localization by looking for general traces of manipulation have also been proposed, see for instance \cite{zhou2018learning,wu2019mantra,marra2020full}.

\enlargethispage{\baselineskip}

\section{Problem statement} 
\label{sec.problemstatement}



\begin{figure}[t]
\centering{\includegraphics[width =0.9\columnwidth]{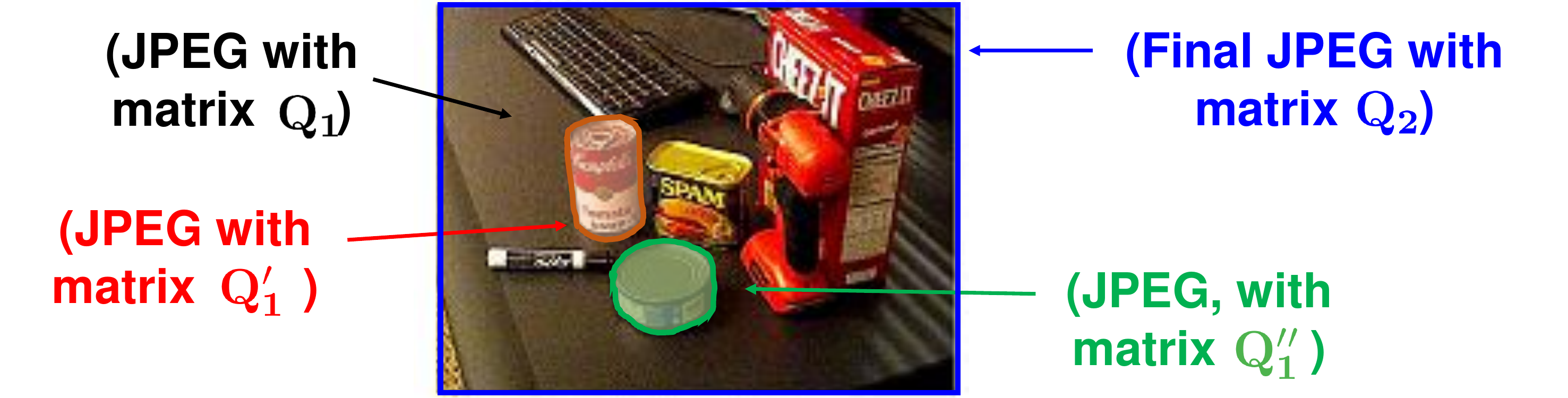}}
\caption{Image tampering setup considered in this paper.}
\label{fig:setup}
\end{figure}

Let $Q$ denote the $8 \times 8$ matrix with the quantization steps of
the DCT coefficients, namely, the quantization matrix, used for JPEG compression.
The image tampering scenario considered in this paper, i.e. the DJPEG vs DJPEG scenario, is illustrated in Fig. \ref{fig:setup}.
The spliced regions (referred to as foreground regions), possibly coming from different  donor images, and the background, are {\em doubly} JPEG compressed. Different quantization matrices have been used for the former compression ($Q_1$). In the figure, $Q_1$ denotes the quantization matrix of the background, $Q_1'$ and $Q_1''$ the quantization matrices of the spliced regions.
The tampered image is finally JPEG compressed with another quantization matrix ($Q_2$). The resulting image is then a double compressed JPEG image.
%
%
%
The second compression can be either aligned or non-aligned to the first one, depending on the position of the $8 \times 8$ JPEG compression grid. A misalignment occurs in the background, for instance, when the image is cropped between the former and the second compression stage. With regard to the spliced area(s), when a region of a JPEG image is copy-pasted into another JPEG image, it is very likely that the alignment between the compression grids is not preserved and the final JPEG compression will not be aligned with the grid of the spliced area(s). In the following, we denote the aligned DJPEG scenario, i.e., when no misalignment occurs between the two compressions, with the acronym A-DJPEG, and the non-aligned DJPEG scenario with NA-DJPEG.

In the scenario described above, spliced regions coming from different donor images can be distinguished by relying on the inconsistencies between the primary quantization matrices. Let $k$ be the total number of donor images. Accordingly, for a pristine image, $k = 1$.  For a tampered image, $k$ corresponds to the number of donor images plus the background and the total number of donor images used to
create the forgery being then $k-1$.
In this setting, the system we have developed aims at solving three different problems: tampering detection, localization and attribution. The detection part outputs a binary decision on the presence or absence of tampering based on the estimated $k$. When tampering is detected ($k>1$), a tampering localization map is returned by the system. The tampering map is a coloured map with different colours assigned to the background pixels and to the pixels of spliced regions coming from different donors. Source
attribution is performed based on the colours of the spliced regions.

In the following, we introduce the main notations used throughout the paper.
%
We denote by ${\bf q}_1$ the 64-dim vector with the elements of the $8 \times 8$ $Q_1$ matrix, taken in zig-zag order \cite{pennebaker1992jpeg}.
The quantization steps corresponding to the medium-high frequencies are more difficult to estimate, since they are quantized more heavily. However, they are usually less important
since they tend to be similar for most quantization matrices. For this reason,
as in most part of related literature, the estimation is restricted to the first $N_c$ elements of ${\bf q}_1$.
Hereafter, we will use the symbol ${\bf q}_{1}$ to indicate only the first $N_c$ quantization steps.
%
With a slight abuse of notation, we denote with $\widehat{Q}_1(\cdot,\cdot,\cdot)$
the tensor with the estimated primary quantization steps computed on $8\times 8$ blocks.
Specifically, for a given $(i,j)$, $\widehat{Q}_1(i,j,v)$ corresponds to the estimation of the $v$-th element of ${\bf q}_{1}$ for the $8 \times 8$ block of pixels in the position indicated by $(i,j)$.

\section{Proposed method}
\label{sec.propmethod}

\begin{figure}[t]
\centering{\includegraphics[width =\columnwidth]{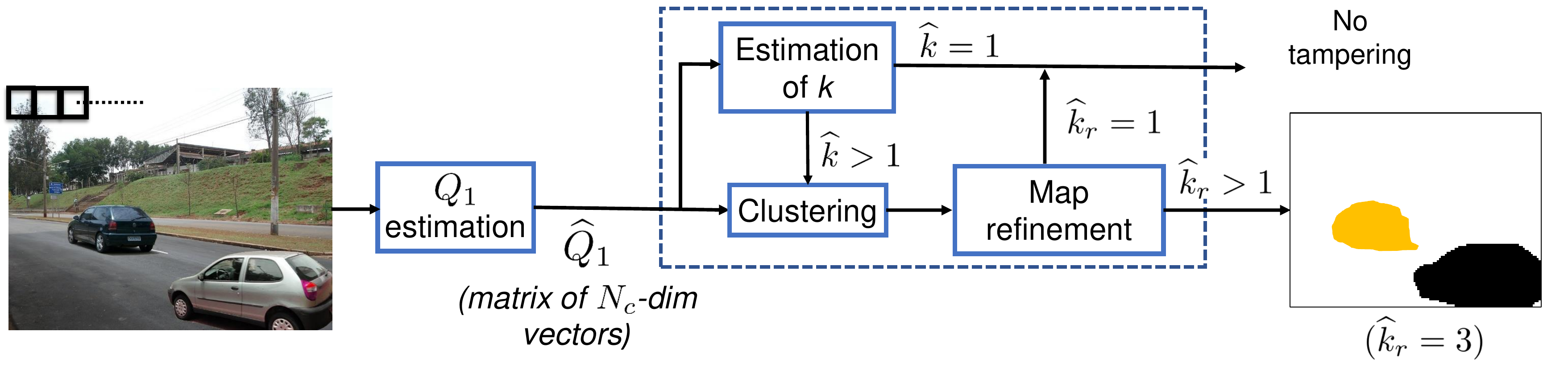}}
\caption{Architecture of the proposed method for splicing detection, localization and attribution of DJPEG images based on $Q_1$ matrix inconsistencies.}
\vspace{-0.5cm}
\label{fig:scheme}
\end{figure}

The overall architecture of the proposed method is illustrated in Fig. \ref{fig:scheme}.
Given a possibly tampered DJPEG image, a block-wise estimation of the primary quantization matrix
is first obtained (first block in Fig. \ref{fig:scheme}), yielding $\widehat{Q}_1$, then splicing detection, localization and attribution is performed by clustering the image blocks according to the result of the estimation followed by a map refinement step (dashed block in Fig. \ref{fig:scheme})\footnote{In principle, a similar pipeline can be used in conjunction with other local features extracted from the image.}.

The number of clusters is first estimated from the $\widehat{Q}_1$  tensor via a CNN model, then a clustering algorithm is applied to $\widehat{Q}_1$   to obtain the tampering map. Specifically, the $N_c$-dim vectors with the estimated quantization steps of the image blocks are regarded as points in the clustering space. As a result of clustering, a label is associated to each block of the image.
In this way, the clustering labels in the tampering map indicate the different donor images used to build the tampered image.
A map refinement step is finally performed on the clustering map to improve the quality of the map based on spatial information.

In the above scheme,  $\hat{k}$ denotes the estimated value of $k$.
Tampering detection is carried out on the basis of the estimated value of $k$.
In particular, if $\hat{k} = 1$, the image is declared to be pristine and the process ends. If $\hat{k} > 1$,
the clustering algorithm is applied to perform splicing localization and attribution, followed by map refinement.
After map refinement, the number of clusters in the map might change. We let $\hat{k}_r$ be the number of clusters after the refinement step. Then, if $\hat{k}_r = 1$, the image is declared to be pristine, while if $\hat{k}_r > 1$,
the image is judged to be tampered.
By
looking at the labels of the different clusters, we can
localize the spliced areas and attribute them to different
donor images.


A detailed description of each block of Fig. \ref{fig:scheme} is provided in the following subsections.

\subsection{Patch-based estimation of the primary quantization matrix}
\label{sec.Q1map}

The primary quantization matrix is estimated on a local window basis, by means of a CNN estimator.
In particular, we chose the CNN-based approach described in \cite{niu2019SPL}, due to its ability to work regardless of the alignment/misalignment of the first and second compression grids and to the good performance obtained even when $QF_2 < QF_1$.

%
%
The network adopted in  \cite{niu2019SPL}  for primary quantization matrix estimation has an input size of $64 \times 64$, and $N_c$ final output nodes, each providing the estimated value of the quantization step of a DCT coefficient.
%
Given an input  patch ${\bf x}$,
the CNN is trained to minimize the difference between the
predicted values ${\bf f}({\bf x})$ and the true vector ${\bf q}_{1}({\bf x})$.
%
%
Rounding is performed  independently on each element of the output vector to get the final prediction, that is ${\hat{\bf q}}_{1}({\bf x}) = \text{round}({\bf f}({\bf x}))$.

Since the CNN is applied patch-wise to the input image, the estimation step returns a tensor with the $N_c$-dim vectors with the primary quantization steps of each $64\times 64$ block.
Specifically, given an input image of size $R \times C \times 3$, the CNN estimator is run on $64\times 64$ overlapping patches (each shifted by 8 pixels with respect to the previous one). The estimated vector, then, is assigned to the central $8 \times 8$ block\footnote{Given that the patch contains an even number of blocks, rigorously speaking a central block does not exist. In the following we denote the block in the fourth block-row and fourth block-column as the central block.} of the patch. At the end, a tensor $\widehat{Q}_1$ with the estimated quantization steps is obtained.
More precisely, by setting to 8 the stride $s$ used to slide the estimation window over the image, and by assuming, for simplicity, that $R$ and $C$ are multiple of 8, the estimated tensor $\widehat{Q}_1$ has size $R' \times C' \times N_c$, where $R' = R/8 -7, C' = C/8 -7$. With reference to the notation introduced in the previous section, $i \in \{1,2,\cdots,R'\}$ and $j \in \{1,2,\cdots C'\}$.
As we said, we set $N_c = 15$.
Note that, for simplicity, we are not considering blocks  close to the border of the image\footnote{If needed, we can incorporate such blocks into the analysis by mirror-padding the border blocks.}.
In the following, we use the compact notation  $\widehat{Q}_{1,t}$ to denote the $t$-th estimated vector,
that is  $\widehat{Q}_{1,t} = {\hat{\bf q}}_{1}({\bf x}_{t})$, where ${\bf x}_{t}$ is the $t$-th $64\times64$ patch fed into the CNN estimator in left-to-right, top-to-bottom, scanning.

\begin{figure}[t]
\centering{\includegraphics[width = 8.8cm]{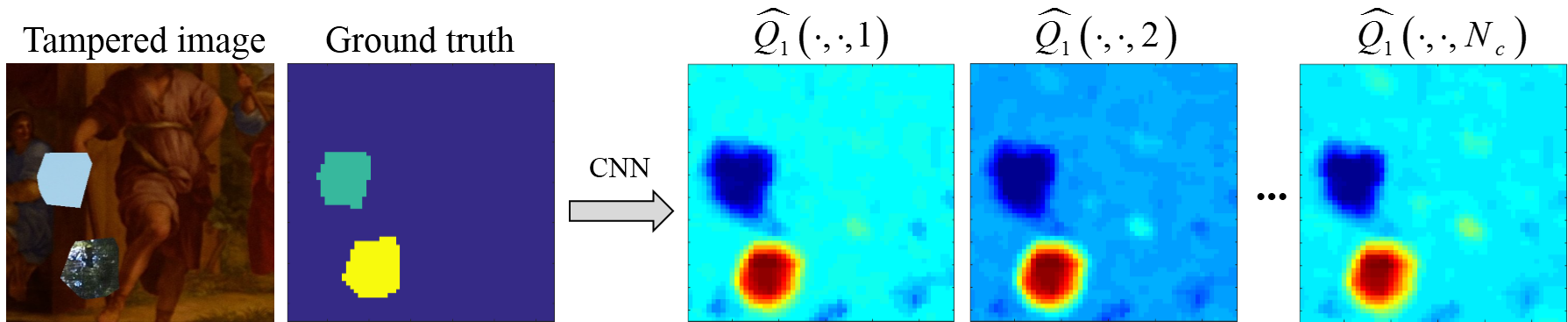}}
\caption{Example of $\widehat{Q}_1$ tensor for various quantization steps. The location of spliced regions can be easily spot from all the bands of the tensor. }
\label{fig:channel}
\vspace{-0.5cm}
\end{figure}

Fig. \ref{fig:channel} shows an example of a tensor $\widehat{Q}_1$ estimated by the CNN. We can observe that the various components of the tensor corresponding to different quantization steps provide useful information regarding the position and the provenance of the spliced areas.

\subsection{Localization and attribution of spliced regions}
\label{sec.clustering_method}

To localize and attribute multiple spliced regions to different donor images based on the estimated quantization matrix, we adopted the Spectral Clustering (SC) algorithm \cite{von2007tutorial}.
SC has been used  in the last years in related image forensics fields, including, for instance, camera identification \cite{amerini2014blind} and mobile phone clustering \cite{li2017mobile}.
Recently, SC has also been considered for splicing detection \cite{mayer2019exposing}, to improve the performance of a deep-learning-based forensic approach for general forgery localization.

Based on some preliminary tests we carried out, SC provides better results compared to other clustering methods, like expectation maximization \cite{mclachlan1988mixture}, hierarchical clustering \cite{Vesanto2000}, fuzzy clustering \cite{1998fuzzy} and, in particular, $K$-means clustering.

\begin{figure}[t]
\centering{\includegraphics[width =0.6\columnwidth]{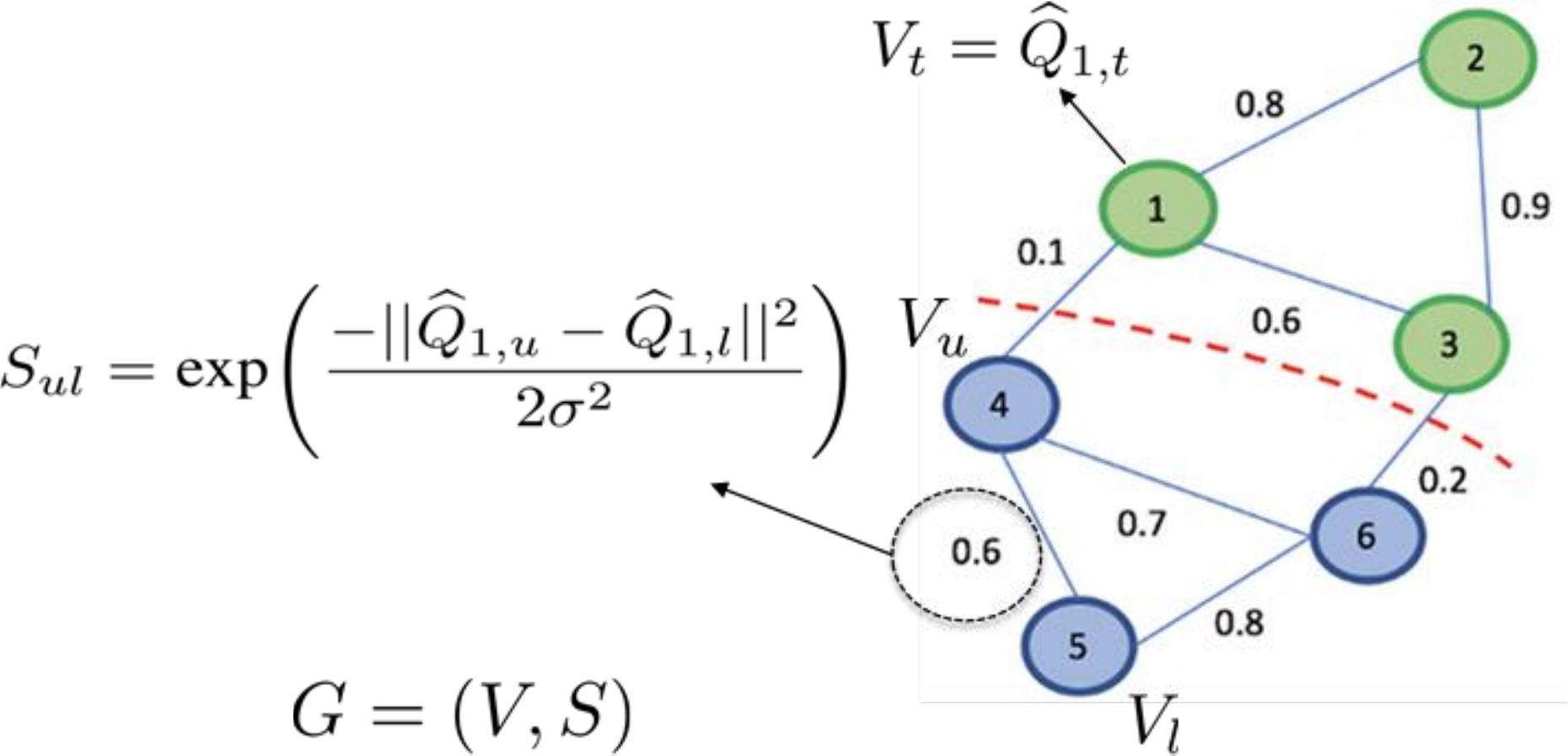}}
\caption{Similarity graph associated to the $\widehat{Q}_1$ tensor.}
\label{fig:graph}
\vspace{-0.2cm}
\end{figure}

SC exploits graph theory to map points (in our case, the $N_c$-dim vectors $\widehat{Q}_{1,t} = {\hat{\bf q}}_{1}({\bf x}_{t})$) to a low dimensional space \cite{von2007tutorial}.
The problem of clustering is reformulated by using a similarity graph.
More specifically, an undirected similarity graph $G = (V, S)$, where $V$ denotes the set of vertexes or nodes
and $S$ the set of edges, is associated to the $\widehat{Q}_1$ tensor as shown in Fig. \ref{fig:graph}.
Each element of the tensor, i.e., each $\widehat{Q}_{1,t}$ vector, represents a node of the graph.

\begin{figure}[!htbp]
\centering{\includegraphics[width = 8.5cm]{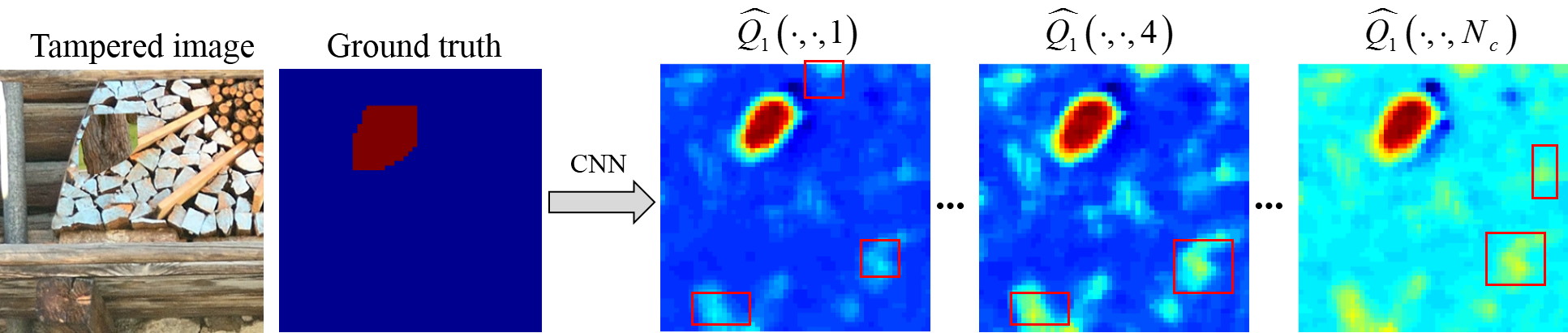}}
\caption{Scattered clusters of $\widehat{Q}_1$ in the spatial domain.}
\label{fig:spatialinformation}
\end{figure}
The number of nodes $N = |V|$ of the graph corresponds to the number of $8 \times 8$ blocks in the image. Then, $S \in \mathds{R}^N \times \mathds{R}^N$. The edge weights $S_{ij}$ represent the similarity of the nodes $i$ and $j$, and in our case they are defined as: $S_{ij} = \exp \left(- ||\widehat{Q}_{1,i} - \widehat{Q}_{1,j}||^2/2\sigma^2\right)$. The choice of the scale parameter $\sigma^2$ is not obvious, so we determined it experimentally.

The goal of SC is to find a partition of the graph such that the edges within a group (cluster) have high weights, i.e., the points within the same cluster are similar to each other, and the edges between different groups (clusters) have very low weights, i.e., the points belonging to different clusters are different from each other.
%
To do so, the SC algorithm computes the Laplacian matrix associated to $G$, then it applies the $K$-means algorithm to the eigenvalues matrix.
%

%

\subsection{CNN-based estimation of the number of clusters}



In principle, the spectral clustering algorithm would provide a way to estimate the number of clusters $k$ by relying on graph theory, that is, through the analysis of the eigenvalues $(\lambda_i)_{i=1}^{N}$ of the Laplacian matrix  associated to the graph (see  \cite{von2007tutorial}). Specifically, the eigenvalues are listed in descending order and the index $i$ corresponding to the maximum gap between two consecutive eigenvalues $\lambda_{i+1} - \lambda_i$ is selected as $\hat{k}$, that is $\hat{k} = \arg\max_i (\lambda_{i+1} - \lambda_i)$.
Based on our experiments, estimating $k$ in this way provides poor results.



By carefully analyzing the results of the preliminary experiments we run to estimate $k$ by means of SC, we concluded that the bad performance we got are due to the fact that SC does not exploit any spatial information.
This prevents to estimate $k$ correctly even in cases that appear easy to solve by visual inspection. 
Fig. \ref{fig:spatialinformation} shows an example in which there are several scattered areas with yellow color in the  estimated tensor $\widehat{Q}_1$. Such scattered areas will be mistakenly regarded as one cluster by SC, even if their spatial incoherence makes it very unlikely that they correspond to a spliced region.  As a result, the estimated $k$ by SC is 3 while the true one is 2.
By exploiting the spatial information, such spatially scattered areas can be identified as a noisy cluster and assigned to the background.

\begin{figure}[t]
\centering{\includegraphics[width =0.8\columnwidth]{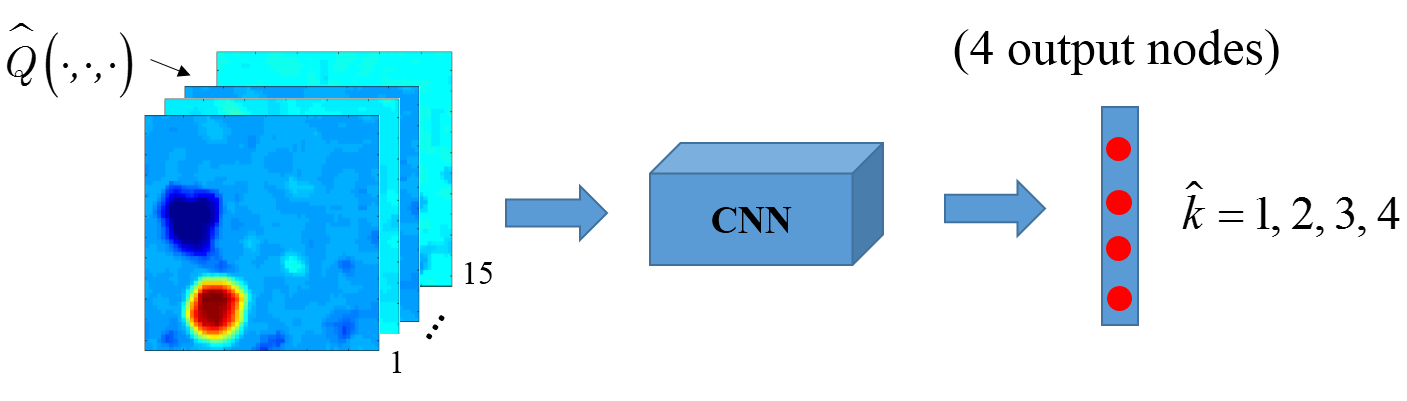}}
\caption{Input and output of the CNN used for estimating $k$ from the $\widehat{Q}_1$ tensor.}
\label{fig:CNN_Kestimation}
\vspace{-0.2cm}
\end{figure}

To do so, we trained a CNN to estimate the number of clusters $k$ directly from the tensor $\widehat{Q}_1$.
%
The CNN has a number of output nodes equal to $4$ (Fig. \ref{fig:CNN_Kestimation}), corresponding to a maximum of 4 clusters \footnote{This choice was made based on the resolution capability of the  estimation of the $Q_1$ matrix. From our experiments, when there are 5 or more distinct donor images with different compression factors $QF_1$s, using a lower number of clusters  yields better localization results. Specifically, the best performance are obtained using $\hat{k} = 4$ (see Section \ref{sec.res_kestimation}).  In these cases, in fact, the difference between the $QF_1$ values is often small (5 or less), then the  estimated coefficients are very close to each other and the clustering algorithm is not capable to correctly separate  the clusters.}.
%
We chose an architecture commonly and successfully used for pattern recognition applications, namely the VGG-16 \cite{simonyan2014very} network. The VGG-16 network has a number of 16 layers in total and 3 Fully Connected (FC) layers.
The input layer is modified to accept input images with size $256 \times 256 \times 15$ (given an input with generic size  $R'\times C' \times 15$, each of the 15-dim quantization maps are resized to normalize the input to the size required by the network).
The CNN has been trained on $\widehat{Q}_1$ tensors computed from pristine images, for $k=1$, and tampered images, for $k = 2,3,4$.
The dataset creation process for the tampered images and the setup considered for network training and validation is described in Section \ref{sec.DB_kestimation},
while the details of the training process are provided in Section \ref{sec.setting}.

\begin{figure}[t]
\centering{\includegraphics[width =8cm]{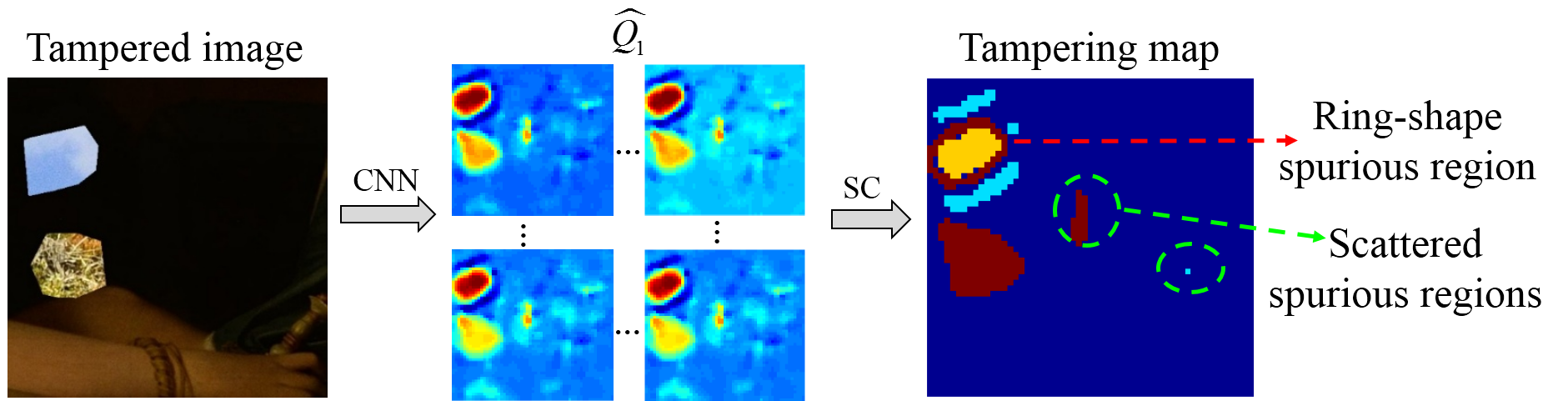}}
\caption{Ring-shape and scattered spurious regions in the preliminary tampering map obtained after clustering.}
\label{fig:MR}
\end{figure}

\subsection{Tampering map refinement by means of morphological reconstruction}
\label{sec.MR}

A visual analysis of the tampering maps obtained after the application of the SC algorithm often reveals the presence of spurious isolated regions that do not correspond to any spliced region. By referring to Fig. \ref{fig:MR} as an example, we observe two types of spurious regions: small and usually scattered regions belonging to the same cluster of one big region, and ring-shaped regions along the boundary of (usually big) spliced areas. The first type of spurious regions are due to the scattered presence of image blocks for which the estimated quantization steps are similar to those of truly spliced regions. The presence of such regions is due to the noisiness of the $Q_1$ estimation step and to the lack of spatial information during the clustering phase. Ring-shaped spurious regions are due to the window-based approach used for $Q_1$ estimation. In the proximity of the boundary of spliced regions, in fact,  the estimation window includes blocks from both the spliced and background regions, thus producing a somewhat mixed estimated vector. For large spliced regions, the number of blocks with the mixed estimate is large enough to represent a separated ring-shaped cluster (the mixed estimated quantization steps may also resemble the values of another truly spliced region, as the brown ring-shaped region in Fig. \ref{fig:MR}). We also observed that spurious regions are more frequent when $\hat{k} > k$, that is when the SC algorithm is run with a  $\hat{k}$ larger than the correct one.

Of course, the presence of spurious clusters reduces the performance of our algorithm in terms of localization (and attribution) accuracy.
For this reason, we introduced a map refinement step, aiming at improving the quality of the tampering map based on spatial information. We did so by resorting to  morphological reconstruction  (MR \cite{gonzales2002digital}).
In particular, the
map refinement procedure consists of the following sequence of morphological operations:
%
\begin{enumerate}
  \item for every cluster we consider the region formed by the pixels belonging to the cluster;
  \item we apply a predefined number of  {\em erosion} iterations with a small structuring element;
  \item we apply a {\em conditional dilation} procedure starting from the regions (referred to as marks or {\em seeds} according to the terminology of morphological reconstruction theory \cite{gonzales2002digital}) obtained at the end of the erosion phase in 2.
\end{enumerate}
%

The conditional dilation procedure works as follows:
i) first, each seed is expanded by means of a conditional dilation, where the dilation is applied only to the pixels belonging to the cluster the region corresponds to. This procedure is carried out in parallel on the seeds of all clusters;
ii) the regions obtained at the end of the previous step, are further dilated conditioned to all the pixels (if any) that do not belong to the background cluster and that have not be assigned yet; iii) if, at a given iteration, regions belonging to different clusters are expanded on the same pixels, disputed pixels are assigned randomly to one of the clusters.

\begin{figure}[t]
\centering{\includegraphics[width = 6.5cm]{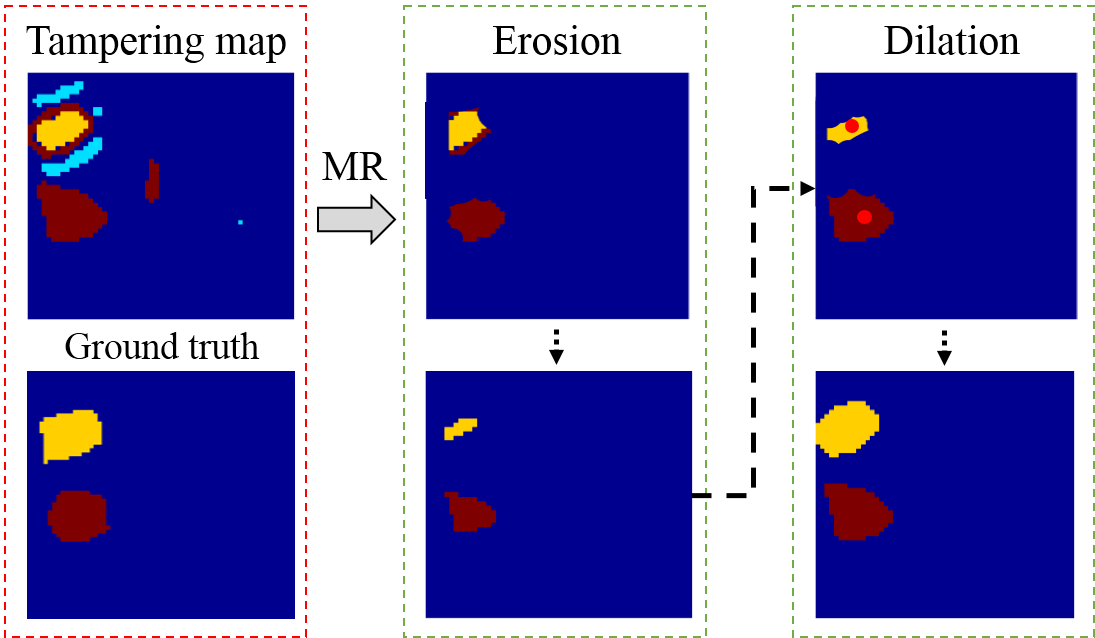}}
\caption{The sequence of morphological operations in MR.
}
\label{fig:MRO}
\vspace{-0.2cm}
\end{figure}

The main goal of MR is to reassign pixels belonging to ring-shaped clusters. Fig. \ref{fig:MRO} shows the results of the process after each of the steps describe above. In the erosion step, the ring-shaped cluster and the isolated cluster are removed while the interior of the spliced regions are kept (see the erosion map). After the conditional dilation procedure, the pixels of the ring-shaped region are reassigned to the corresponding inner cluster (see the dilation map).
Note that isolated (usually small) regions are completely eroded during the erosion step and are not reassigned during the conditional dilation.
The choice of the number of erosion iterations and the size of the structuring element  is a crucial one, all the more that the optimum setting depends on the size of the spliced areas.
For a given structuring element, a too large number of iterations may cause the removal of spliced regions, while if the number of iterations is too small, the risk is that small isolated clusters are not removed.
Given the difficulties of determining the best setting on a theoretical basis, we tuned the system by means of experimental analysis.

\begin{figure}[t]
\centering{\includegraphics[width =0.8\columnwidth]{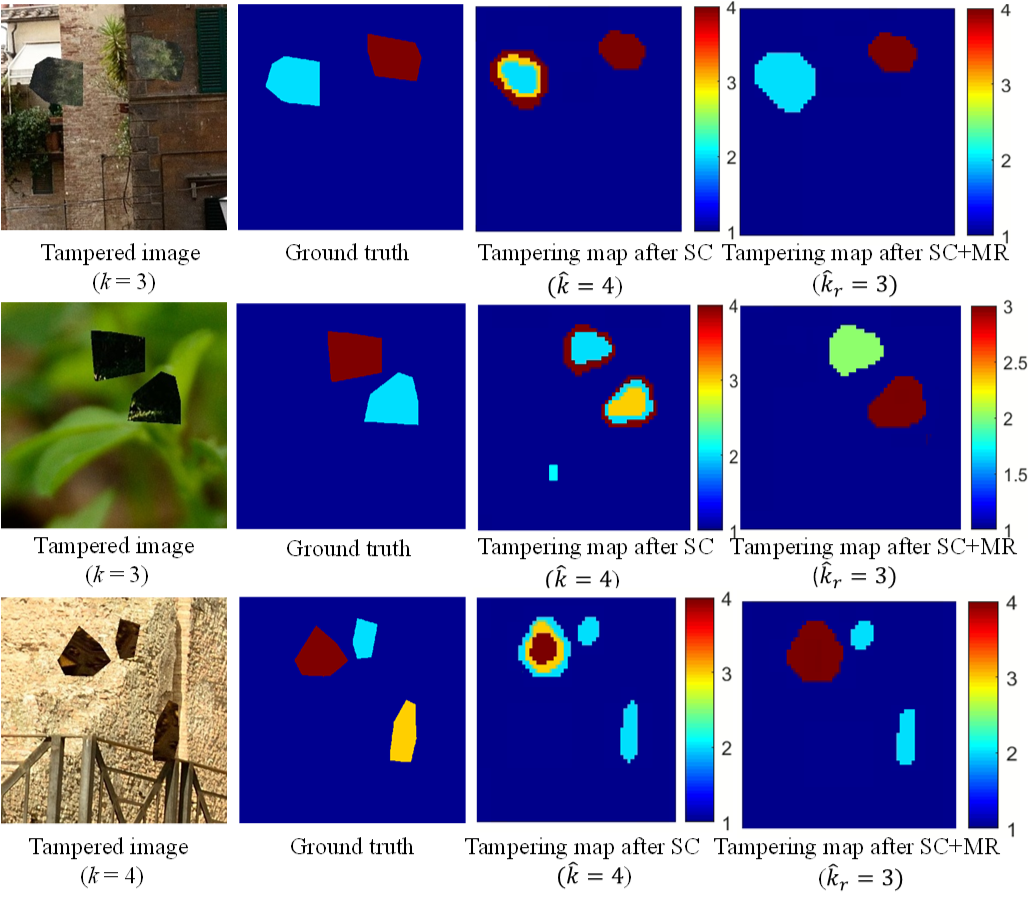}}
\caption{Examples of tampering maps before and after the application of morphological reconstruction.}
\label{fig:MRexamples}
\vspace{-0.5cm}
\end{figure}

After the application of the MR procedure, the number of clusters might change. In particular, the final number $\hat{k}_r$ of clusters may be {\em lower} than $\hat{k}$,
because isolated clusters have been removed or ring-shaped clusters have been reassigned to the internal clusters.
This happens especially when $k \ge 3$ (see Section \ref{sec.res_kestimation}). If after MR the number of remaining clusters is equal to 1, the image is considered to be a pristine one.
Experiments carried out on several tampered images reveal that MR can indeed help to remove noisy clusters and the undesired rings around compact regions from the maps.
The benefits obtained with the map refinement procedure are illustrated in the examples reported in Fig. \ref{fig:MRexamples}.
The number of erosion iterations considered in those examples is 2 and the size of the (disk-shaped) structuring element is 1.
In the examples reported in the figure, at the end of the MR procedure,
the ring-shaped clusters are reassigned to the internal clusters and then $\hat{k}_r < \hat{k}$.
We notice that the estimation of $k$ can be worse after map refinement, however, especially when $k$ is large, a better clustering result is obtained when the value of $k$ is underestimated. This is the case with the last example in Fig. \ref{fig:MRexamples}, where we see that using $3$ clusters instead of $4$ permits to remove the  two rings around one of the spliced regions. This, apparently counterintuitive, behavior is due to the fact that, especially when $k$ is large, the value of the quantization coefficients may be similar for different donor images, and hence it might not be easy to correctly identify the clusters using the true $k$. In such cases, a better map is obtained by assigning the regions originating from the donor images with similar quantization matrices to the same cluster.

\section{Experimental Methodology}
\label{sec.methodology}


In this section, we
describe the methodology we followed to run the
experiments whereby we validated the effectiveness of the
proposed method.
We first present the evaluation metrics used to measure the performance of the proposed tool to detect and localize the tampered areas. Then we introduce a metric explicitly thought to measure the effectiveness with regard to the attribution task.
Afterwards, we pass to the description of the procedure that we
followed to generate the tampered contents in the DJPEG scenario, and introduce the datasets used for: i) training and testing the CNN model for the estimation of $k$; ii) assessing the detection, localization and attribution performance of the system.
Finally, we describe the two closest related state-of-the-art methods
and describe how they are applied for a fair comparison with the results obtained by our system.

For simplicity and without loss of generality, in most of our experiments, we
considered standard quantization matrices, so we refer to them by means of the Quality Factor ($QF$).
Specifically, we denote with $QF_2$ the
$QF$ used for the second compression and with $QF_1$ that of the primary compression.

\subsection{Evaluation metrics}
\label{sec.eval_metrics}


In this section, we introduce the metrics used to measure the performance of our algorithm.

\subsubsection{Tampering detection} 

Tampering detection is a binary classification problem.
For tampered images, we have a correct detection when $\hat{k}_r > 1$, while for pristine images, the detection is correct when $\hat{k}_r = 1$, wrong in all the other cases.

By adopting a common terminology in detection theory, in the following we assume that tampered images ($k > 1$) belong to the positive class and pristine images ($k = 1$) to the negative one.
A Neyman Pearson setup is considered for the decision.
Accordingly, we fixed the maximum admissible
False Positive Rate (FPR = $Pr(\hat{k}_r > 1|k = 1)$), i.e., the percentage of pristine images wrongly detected as tampered, and we evaluate the True Positive Rate (TPR = $Pr(\hat{k}_r > 1|k > 1)$), namely, the percentage  of correctly detected tampered images.
The overall accuracy is given by the fraction of correct decisions on both tampered and pristine images over the total number of tested images.

\subsubsection{Tampering localization and attribution}
\label{sec.metric_NMI}

With regard to the metrics for assessing the localization and attribution performance, we observe that
we should not only evaluate the capability of the system to localize the tampered areas,
but also the capability to identify the regions spliced from different donor images as belonging to different clusters.

Let us first focus on the former task. Tampering localization can be regarded as a binary classification problem applied at the pixel level. Pixels belong to one of two classes, the background (the negative class) or the foreground (tampered or positive class).
To measure the localization performance, we consider the Matthews Correlation Coefficient (MCC) \cite{matthews1975}, defined as:
%
\begin{equation}
\text{MCC} = \frac{\text{TP} \times \text{TN} + \text{FP}\times \text{FN}}{\sqrt{(\text{TP}+\text{FP})(\text{TP}+\text{FN})(\text{TN}+\text{FP})(\text{TN}+\text{FN})}},
\label{MCC}
\end{equation}
%
%
where TP is the number of true positives pixels, i.e., the pixels correctly classified as tampered, TN the number of true negative pixels, i.e., the pixels correctly classified as non-tampered, FP the number of false positives and FN the number of false negatives. If any of the sums in brackets at the denominator is zero, the denominator
is arbitrarily set to one.
MCC is particularly helpfull in the case of unbalanced classes, as it is almost always the case for tampering localization\footnote{In the case of highly unbalanced classes, the overall accuracy is not a good indicator of the performance given that errors on the minority class have virtually no impact.}.

The identification of spliced regions coming from different donor images is a new goal addressed in this paper, so no established metric exists to measure the performance with respect to this task. To fill this gap, we introduce a new metric called Normalized Mutual Information (NMI) \cite{dataMining}.

To start with, we observe that the output of the clustering algorithm is a map assigning to each pixel a label, ranging from $1$ to $\hat{k}_r$, indicating the cluster the pixel belongs to. The ground-truth map indicates for every pixel the true cluster, i.e., the corresponding donor image for foreground pixels or the background cluster. Note that, the exact label assigned to each cluster is irrelevant, as long as pixels coming from differ donor images are assigned to different clusters. So it is not necessary that the labels of the clustering map are identical to those of the ground truth. As an additional difficulty, we observe that the number of clusters contained in the map output by the tampering localization and attribution system does not need to be equal to the number of clusters in the ground-truth map.

In the following, we will refer to the labels assigned to the regions of the ground-truth map as pixel classes.
Let  $y$ denote  the class label ($y = 1,..,k)$ and $c$ denote the cluster label ($c=1,...,\hat{k}_r$) in the output clustering map.
The NMI index is defined as:
\begin{equation}
\text{NMI}(y,c) = \frac{2 \text{I}(y;c)}{\text{H}(y) + \text{H}(c)},
\end{equation}
where H$(y)$ and H$(c)$ denote, respectively, the empirical Entropy of $y$ and $c$, and I$(y;c)$  the empirical Mutual Information between $y$ and $c$.
More precisely, let $p_y(i) =  p(y = i) = \#\text{\em \{pixels in class $i$\}}/\text{\em total no. of pixels}$, $p_c(j) =  p(c = j) = \#\text{\em \{pixels in cluster $j$\}}/\text{\em total no. of pixels}$, and $p_{y|c}(i | j)  = p_{y|c}(y = i| c=j)$ where
\begin{align}
p_{y|c}(i | j) 
=\frac{\#\text{\em \{pixels of class $i$ assigned to cluster $j$\}}} {\#\text{\em \{pixels in cluster $j$\}} },
\end{align}
%
%
%
where the total number of pixels is $R'\cdot C'$.
Then,
%
$H(y) = \sum_{i=1}^{k} p_y(i) \log p_y(i)$,
%
and
\begin{equation}
I(y;c) = \sum_{i=1}^{k} \sum_{j = 1}^{\hat{k}} p_{y|c}(i|j) p_c(j) \log \Big(\frac{p_{y|c}(i|j)}{p_y(i)}\Big).
\end{equation}

In case of perfect clustering, it is easy to see that $H(y) = H(c)$, while $I(y;c) = H(y)$, then NMI = 1.
Note that being a normalized quantity, the NMI allows comparing cases with a different number of clusters.


\enlargethispage{\baselineskip}

\subsection{Datasets}
\label{sec.dataset}



To build the datasets for our experiments, we started from the 8156 camera-native uncompressed large size images in the RAISE8K dataset \cite{RAISE8K}. We divided these images in two sets: 7000 images to be used for training (and validation) and 1156 images for the tests.
%
On the average, about 5 non-overlapping patches
are extracted from each RAISE image, for a total number of 41000 patches, 35000 of which (coming from the set of 7000 original images), denoted by $\mathcal{S}_{tr}$, were used to produce the pristine and tampered images for training the models and the remaining 5780, denoted by $\mathcal{S}_{ts}$, to produce the pristine and tampered images for the tests.

%
We considered two types of DJPEG pristine and tampered images, named Type I and Type II, described below,
respectively for the case of Aligned DJPEG (A-DJPEG) and the case of  Non-Aligned DJPEG (NA-DJPEG)\footnote{To be precise, we refer to the Type I (Type II)
set as  aligned (not-aligned) set by implicitly referring to the compression
of the pristine images in the set, and, similarly, to the compression of the background of the tampered regions, while the compression of the foreground is always misaligned.}. 
\begin{itemize}
\item For {\bf Type I} images:  the pristine images are A-DJPEG images. This means that the first 8$\times$8 DCT compression grid is aligned with the grid of the second compression. With regard to the tampered images, the first JPEG compression grid of the background (or, equivalently, of the source image) is aligned with the grid of the second compression, while the grid for the first compression of the foreground is misaligned with that of the second compression.
    Specifically, a random misalignment $(r,c)$, $0 \leq r, c \leq 7, (r,c)\neq (0,0)$ is considered for the grid of the former compression of the foreground.
\item For {\bf Type II} images: we assume that the images are first JPEG compressed using a DCT grid shifted by a quantity $(r,c)$, randomly chosen in $0 \leq r, c \leq 7, (r,c)\neq (0,0)$, with respect to the upper left corner, while for the second compression no grid misalignment is considered. Then, the pristine images are NA-DJPEG. For the tampered images, the JPEG grid of the background is non-aligned with the grid of the second compression. The same holds for the foreground regions. Note that the misalignments of foreground and background are generally different.
\end{itemize}

The datasets we used for our experiments are available online, together with a report detailing the exact procedure we have followed to build the pristine and tampered images for the various $k$ and combination of $QF_s$\footnote{The document is made available at:
\url{https://drive.google.com/drive/folders/1ck-Xm1G3dxgGN717B_JVdKMpBaPCZ3Ap}, along with the datasets.}. Below we provide a description of the datasets considered for the various tests. The size of the images in all the datasets is 512$\times$512.

\subsubsection{Dataset for $Q_1$ matrix estimation (training and testing)}
\label{sec.DB_kestimation}

To build the datasets for training and testing the CNN for the estimation of $Q_1$, we followed exactly \cite{niu2019SPL}.
Training and testing were carried out on 64$\times$64 patches, obtained from the set of 7000 and 1156 images of RAISE. For DJPEG, a random grid shift $(r,c)$ is applied between the two compressions, $0 \leq r, c \leq 7$, then, as in \cite{niu2019SPL}, the A-DJPEG case occurs with probability 1/64.

\subsubsection{Dataset for $k$ estimation (training and testing)}

The dataset used to train and test the CNN for the estimation of the number of clusters consists of:
%
\begin{itemize}
\item a set $\mathcal{D}_{tr}$ of 18000 images for each $k$ (for a total of 72000 images)  used for training. The set is obtained from 18000 (randomly chosen) images in $\mathcal{S}_{tr}$;
\item a set $\mathcal{D}_{ts}$ of 4000 images for $k=1$  and  4000 images for $k>1$, in equal proportions for $k=2,3$ and $4$. The set is obtained from 4000 images in $\mathcal{S}_{ts}$.
\end{itemize}

Let $V = \{60,65,70,75,80,85,95,98\}$. To get the pristine images, $QF_1$ is randomly chosen in $V$ and $QF_2 = 90$.
For the tampered images, when $k=2$, $QF_{1}$ is randomly chosen in $\{75,85,95,98\}$, and $QF_{1,2} \in V$, $QF_{1,2} \neq QF_{1}$. When $k=3$, $QF_{1}\in \{75,85,95,98\}$, and $QF_{1,2}, QF_{1,3} \in V$, $QF_{1,2} \neq QF_{1,3}  \neq QF_{1}$. Finally, for $k=4$, $QF_{1}\in \{75,85,95,98\}$, and $QF_{1,2} \ne QF_{1,3} \ne QF_{1,4} \ne QF_1 \in V$.
%
The height $h$ and width $w$ of the bounding-box of the tampered regions are randomly selected in \{64, 96, 128, 156\}.
Misalignment is applied to the background with 0.5 probability, then the dataset consists of both Type I and Type II images in similar proportions.


\subsubsection{Dataset for detection, localization and attribution tests}

%

Detection performance are measured over the same dataset $\mathcal{D}_{ts}$ considered to test the CNN for $k$ estimation, where we have  4000 images representative of the negative class (pristine), and 4000 for the positive class (tampered). The threshold achieving the desired FPR is set on these 4000 pristine images.
%
To better assess the localization performance, and to ease the comparison with state-of-the-art methods (see the next section), we additionally built two separate Type I and Type II datasets, named $\mathcal{D}_I$ and $\mathcal{D}_{II}$,
whose images are generated from  $\mathcal{S}_{ts}$ under specific setting. Specifically, in both $\mathcal{D}_I$ and $\mathcal{D}_{II}$, we considered 100 images for every combination of $k$ and $\{QF_{1,i}\}_{i=2}^k$, for the tampering sizes $h \times w = 96\times 96$ and $128 \times 128$.

A summary of the datasets used in our experiments is reported in Table \ref{tab.datasets}.


\begin{table}
\centering
\setlength{\tabcolsep}{2pt}
\scriptsize
\caption{Datasets of images  considered in our experiments.}
	\vspace{0.1cm}
	{\begin{tabular}{|l||c|c||c|c|}
            \hline
            \textbf{Name} & $\mathbfcal{D}_{\bf tr}$ & $\mathbfcal{D}_{\bf ts}$  & $\mathbfcal{D}_{\bf I}$ &  $\mathbfcal{D}_{\bf II}$ \\ \hline
             {Purpose} & Training  & Test  &  Test  &  Test \\ \hline
            {Original} & \multirow{2}{*}{$\mathcal{S}_{tr}$ } & \multirow{2}{*}{$\mathcal{S}_{ts}$}  &  \multirow{2}{*}{$\mathcal{S}_{ts}$} &  \multirow{2}{*}{$\mathcal{S}_{ts}$} \\
            {dataset} &  &  &   & \\ \hline
            \multirow{2}{*}{No. images} & 18000 per $k$  & 4000 per $ k=1$  &  100 per each   ($k$, & 100 per each   ($k$,  \\
            &  (72000 total) & 4000 per $k>1$  &  $\{\text{QF}_{1,i}\}$,  h $\times$ w) & $\{\text{QF}_{1,i}\}$, h $\times$ w) \\ \hline
            \multirow{2}{*}{DJPEG} & Type I and II & Type I and II &   \multirow{2}{*}{Type I} & \multirow{2}{*}{Type II}  \\
             & (50\% each) & (50\% each) &   &   \\             \hline
            \multirow{3}{*}{Setting} & $\{\text{QF}_{1,i}\}$, h$\times$w   & $\{\text{QF}_{1,i}\}$, h$\times$w  &   h$\times$w  =   & h$\times$w  = \\
            &  randomly  &  randomly &   $\{96 \times 96,$  & $\{ 96 \times 96$ \\
            & chosen & chosen &  $128 \times 128\}$   & $128 \times 128\}$\\ \hline
	\end{tabular}}
	\label{tab.datasets}
\end{table}

\subsubsection{Additional datasets}

Although most of the experiments were carried out using the datasets described so far, to test the generalization performance of our method in different and more challenging scenarios, the following datasets have also been considered:
\begin{itemize}
  \item The datasets used in \cite{bianchi2012image}, named $\mathcal{D}_{\footnotesize \cite{bianchi2012image}}$, whose images have resolution 1024$\times$1024. These images are obtained from a personal database of uncompressed images. $QF_1 \in \{50:100\}$ with step 5. The DJPEG vs SJPEG scenario is considered in this dataset for the tampering. The tampering region corresponds to the central portion (1/16 of the image). Two datasets are provided, one for the  aligned case ($\mathcal{D}_{{\footnotesize \cite{bianchi2012image}},\text{A}}$) - the foreground is A-DJPEG -, and one for the non-aligned case ($\mathcal{D}_{{\footnotesize \cite{bianchi2012image}},\text{NA}}$) -  the foreground is NA-DJPEG. In all cases $k = 2$, in particular, we considered the images with $QF_2 = 90$ for comparison.
  \item The dataset used \cite{wang2014exploring}, named $\mathcal{D}_{\footnotesize \cite{wang2014exploring}}$, whose images have been obtained starting from the uncompressed images in the UCID dataset \cite{UCID}, with size 512$\times$384.  The primary quality factor $QF_1$ was chosen in the set $\{55,65,75,85,95\}$.   The DJPEG vs SJPEG scenario is considered for the tampering. The tampering region is 10\% of the image, whose location is randomly chosen. In all cases $k = 2$. For our comparison, we considered the images with $QF_2 = 90$.
  \item A dataset built as detailed abeve but starting from uncompressed images of Dresden dataset \cite{Dresden}, named $\mathcal{D}_{Dr}$. For the tampered area we let  $h\times v$ = 96 $\times$ 96, 128 $\times$ 128, and 156 $\times$ 156. We built Type I and Type II images with  $k = 2,3,4$. For each setting, we considered 100 images for every combination of $QF_1$.

  \item A dataset built from $\mathcal{S}_{ts}$ in which the first compression was carried out by using Photoshop, with PS qualities in the range [7:12] (the second compression is the same as before). 
The more general non-aligned scenario is considered.
The tampering sizes are set to $h\times v$ = $96 \times 96$, $128 \times 128$, and $156 \times 156$ with $k=2$.  100 images are considered for each PS quality in each setting. The results  are reported in Table \ref{tab.PS}.
\end{itemize}

\begin{figure}[t]
\centering{\includegraphics[width =0.85\columnwidth]{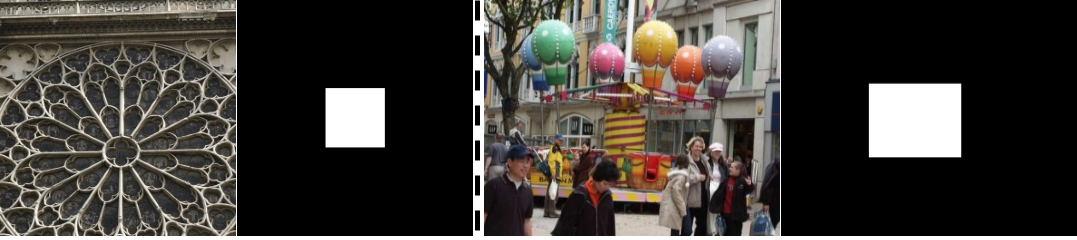}}
\caption{Examples of tampered images from $\mathcal{D}_{\footnotesize \cite{bianchi2012image}}$ (left) and $\mathcal{D}_{\footnotesize \cite{wang2014exploring}}$ (right).}
\label{fig:tamperingBianchiWang}
\end{figure}

Examples of tampered images from $\mathcal{D}_{\footnotesize \cite{bianchi2012image}}$ and $\mathcal{D}_{\footnotesize \cite{wang2014exploring}}$ are provided in Fig. \ref{fig:tamperingBianchiWang}. These images are built by cutting a central portion of the image, compressing it and pasting it back in the same position, thus not leaving any visual boundary artifacts.

\enlargethispage{\baselineskip}

\subsection{Baseline methods for comparison}
\label{sec.method_comparision}


The baselines we compared our method with are the methods described in \cite{bianchi2012image} and \cite{wang2014exploring},  since, as we said in  Section \ref{sec.ArtPrior}, these are the methods most closely related to the system proposed in this paper.
%
In \cite{bianchi2012image}, statistical models are used to build a map reporting the likelihood that image blocks have been double compressed. The cases of A-DJPEG and NA-DJPEG are treated separately. In \cite{wang2014exploring}, the authors propose a
strategy to estimate the posterior probability that an image block has been tampered with, by minimizing a properly defined energy function. The approach works only for the case of A-DJPEG compression.
Similarly to our system, both these methods are designed for tampering localization in double compressed images and can be applied to a scenario wherein both the background and spliced regions are double compressed but exhibit different compression artifacts (i.e., they are compressed with a different quality factor), since they are based on the estimation of the DCT quantization steps used for the first compression.
Both methods perform well when the former compression quality is {\em lower} than the second, while they provide poor performance in the reverse case.

To turn the tampering localization map provided by the baseline methods into a tampering detection output, we followed the approach used in \cite{wang2014exploring} and trained an SVM classifier with 3 features. The first feature is given by the perimeter-area ratio of the localized tampered region. The second one is the percentage of pixels detected as tampered. These two features are used to characterize scattered and small regions, that are usually linked to false alarms.
The third feature is a measure of the periodicity consistency between the DCT coefficient histograms of the localized region and the entire image.
We refer to \cite{wang2014exploring} for more details.
To train the SVM, we considered 3000 pristine ($k=1$)  and 3000 tampered images (for $k=2,3,4$, where each $k$ is represented in equal proportions),
obtained from the images in $\mathcal{S}_{tr}$ as detailed in the previous section, with the difference that only Type I images are considered for \cite{wang2014exploring}, while  Type I images and Type II images are considered for the methods in \cite{bianchi2012image}, respectively for the A-DJPEG and NA-DJPEG cases\footnote{The same proportion of pristine and tampered images is considered in these cases (the SVM is trained for the binary tampering detection task).}.
%
Given the trained model, the operating point is determined from the ROC curve by fixing  the desired FPR and deriving the SVM decision threshold accordingly.
About 2000 pristine images obtained from the images in $\mathcal{S}_{ts}$ (different from those used to build the datasets for the tests) are considered to set the threshold, for the Type I setting and Type II setting respectively.

We stress that the comparison with \cite{bianchi2012image} and \cite{wang2014exploring} is possible only when the considered setting satisfies the operative conditions such methods have been built for. In fact, an advantage of our method is that it is much more general than \cite{bianchi2012image} and \cite{wang2014exploring}, so that it can work in a wider variety of situations. In addition, our method is able to distinguish between spliced regions coming from different donor images, which is something that neither \cite{bianchi2012image} nor \cite{wang2014exploring} can do. Such aspects must be taken into account for a fair judgement of the  improvement allowed by the method proposed in this paper. The operative conditions, and expected performance of the various methods under different settings, are summarized in Table \ref{tab.comparison}.

For sake of completeness, we also compared our method
with an anomaly-based detector that performs forgery localization by looking for general traces of manipulation, hence not focusing on DJPEG compression artifacts. In particular, we considered a recently proposed deep learning method, named MantraNet \cite{wu2019mantra}, based on anomaly detection, that performs joint image-level detection and pixel-level localization of forgeries,  regarded as local image anomalies.
\begin{table}[t]
\centering
\scriptsize
\caption{Operative conditions of our system and the methods in  \cite{bianchi2012image} and  \cite{wang2014exploring}.}
	\vspace{0.1cm}
	{\begin{tabular}{|l|c|c|c|c|c|}
            \cline{2-5}
            \multicolumn{1}{l|}{\multirow{2}{*}{}}  & \multirow{2}{*}{DJPEG grid} & {Background} & Non-std  & \multirow{2}{*}{Purpose} \\
            \multicolumn{1}{l|}{\multirow{2}{*}{}} &  & setting  & $Q_1$ matrix  & \\ \hline
            \multirow{2}{*}{\cite{bianchi2012image}-Al} &  \multirow{2}{*}{Aligned} & $QF_1 < QF_2$,  & \multirow{2}{*}{Yes} & \multirow{2}{*}{Localization} \\
            &   & $QF_1 > QF_2$ (poor)  & &  \\ \hline
                        \multirow{2}{*}{\cite{bianchi2012image}-NAl} &  \multirow{2}{*}{Non-Aligned} & $QF_1 < QF_2$,  & \multirow{2}{*}{Yes} & \multirow{2}{*}{Localization} \\
            &   & $QF_1 > QF_2$ (poor)  & &  \\ \hline
                        \multirow{2}{*}{\cite{wang2014exploring}} &  \multirow{2}{*}{Aligned} & $QF_1 < QF_2$,  & \multirow{2}{*}{Yes} & \multirow{2}{*}{Localization} \\
            &   & $QF_1 > QF_2$ (poor)  & &  \\ \hline
                                    \multirow{2}{*}{Our} &  \multirow{2}{*}{Both} & \multirow{2}{*}{$QF_1 \lessgtr QF_2$}  & \multirow{2}{*}{Yes} & Localization \\
            &   &   & & Attribution \\ \hline
	\end{tabular}}
	\label{tab.comparison}
\end{table}

\subsection{Parameter setting}
\label{sec.setting}

Below, we report the setting of the parameters considered to implement the various steps of our system.
To train the CNN for $Q_1$ estimation, we followed exactly \cite{niu2019SPL}.
Then, the network is trained on $64\times 64$ patches ($4 \times 10^5$ for each $QF_1$) obtained from the set of 7000 RAISE images used for training as detailed in \cite{niu2019SPL}.
The network is trained for 60 epochs with batch size 32.
The solver is the Adam optimizer. The learning rate is $10^{-5}$.
The $\widehat{Q}_1$ map is obtained as described in Section \ref{sec.Q1map}.
%
%
With regard to the CNN used to estimate $k$, as we said, we considered the VGG-16 network \cite{simonyan2014very}. We started from the pre-trained solution for ImageNet classification, and re-trained the network on the dataset described in Section \ref{sec.DB_kestimation} for 50 epochs, with batch size 16 and applying
data augmentation. The Adam optimizer with learning rate $10^{-5}$ was considered as solver.
Regarding the spectral clustering algorithm, the scale parameter $\sigma$ is set to  $0.6$ if the estimated $k$ is $2$, while we use  0.15 if $k = 3$ or $4$.
%
Finally, for the morphological reconstruction, we considered 2 erosion iterations with a disk-shaped structuring element  of radius 1.  We found experimentally that this setting is a good choice for the range of tampering sizes that we are considering.

\section{Experimental results}
\label{sec.results}



The results of our experiments are reported and discussed in this section for a wide variety of settings.

\subsection{Accuracy of $k$ estimation}
\label{sec.res_kestimation}

%
%
%

The estimation of $k$ is a crucial step since it directly affects the performance of tampering detection, localization and attribution. In particular, if $k=1$ and the estimated $\hat{k}_r>1$, pristine images are mistakenly identified as tampered images, and vice versa.

The results of the $k$ estimation step are reported in the confusion matrix shown in Table \ref{tab.ConfusionMatrix} (left).
The average accuracy of the estimation is 0.79. Somewhat expectedly, the estimation works better for smaller $k$, since the minimum difference between the $QF_1$ values considered tends to be smaller in the presence of multiple spliced regions.

Table \ref{tab.ConfusionMatrix} (right) reports the number of clusters $\hat{k}_r$ after spectral clustering and map refinement. The average accuracy of the estimation now is 0.75.
Although $\hat{k}_r$ is a worse estimate, better localization results can be obtained with $\hat{k}_r$. In fact, as we mentioned at the end of Section \ref{sec.MR},  in some cases, better clustering results can be obtained by underestimating $k$. This is especially the case for large $k$, when using the true value may result in wrong clusters, e.g. ring-shaped and noisy clusters   - see the examples in Fig. \ref{fig:MRexamples}. This is confirmed by the results of our  ablation study (end of Section \ref{sec.attribution}).
For the same reason, we also found that there is no need to consider more than 4 clusters: in fact, when $k>4$, the best localization results are obtained using $\hat{k} = 4$, that is, assigning the regions originating from  donor
images with similar quantization matrices to the same cluster. With $k=5$, for instance, we obtain a  gain of 0.007 in MCC and 0.004 in NMI by using 4 clusters instead of 5. The gain is more significant for larger $k$.%

\begin{table}
\centering
\setlength{\tabcolsep}{3pt}
\caption{Confusion matrix of the CNN-based $k$ estimation (left) and of the final $\hat{k}_r$  (right), computed over the $\mathcal{D}_{ts}$  set. 
}
	\vspace{0.1cm}
	{\begin{tabular}{|c|c|c|c|c|}
            \hline
            $k \setminus \hat{k}$ & 1 & 2 & 3 & 4\\ \hline
			1 & {\bf 0.95} & 0.05 & 0.00   & 0.00 \\ \hline
            2 & 0.09 & {\bf 0.82} & 0.08 & 0.01  \\ \hline
            3 & 0.01 & 0.21 & {\bf 0.67} & 0.11 \\ \hline
            4 & 0.00 & 0.04 & 0.24 & {\bf 0.72} \\ \hline
	\end{tabular}}
	{\begin{tabular}{|c|c|c|c|c|}
            \hline
            $k \setminus \hat{k}_r$ & 1 & 2 & 3 & 4\\ \hline
			1 & {\bf 0.95} & 0.05 & 0.00    & 0.00 \\ \hline
            2 & 0.09 & {\bf 0.84} & 0.06 & 0.01  \\ \hline
            3 & 0.01 & 0.28 & {\bf 0.64} & 0.07 \\ \hline
            4 & 0.00 & 0.09 & 0.36 & {\bf 0.55} \\ \hline
	\end{tabular}}
	\label{tab.ConfusionMatrix}
\end{table}
%

%

\subsection{Detection performance ($k=1$ vs $k>1$)}

%

%
With regard to the detection performance ($k = 1$ vs $k > 1$) of our system, it can be derived from Table \ref{tab.ConfusionMatrix} (right) by measuring the capability of distinguishing pristine ($k = 1$) from tampered images ($k > 1$). The detection accuracies we got for $k=1$, $k>1$ and over the $\mathcal{D}_{ts}$ set (we remind that this set comprises both Type I and Type II images in equal percentage) are 0.95, 0.97 (see Table \ref{tab.Accuracy}) and 0.96 respectively;
 specifically, we got  TPR = 0.97  at FPR = 0.05.
Table \ref{tab.DetPerformance} (last row) reports the accuracy results split for different settings.
Specifically, we got TPR = 0.97 for Type I and 0.96 for Type II tamperings.

The comparison with \cite{bianchi2012image} and \cite{wang2014exploring}, where detection is performed via SVM classification as described in Section \ref{sec.method_comparision} is reported in Table \ref{tab.DetPerformance}, where TPR values corresponding to FPR = 0.05 are reported for both methods in Type I and Type II scenarios. For the baselines, the table reports only values obtained under the operative conditions the methods were thought for.
As it can be seen, the proposed method greatly outperforms the baselines in the various settings. The poor results of \cite{bianchi2012image} and \cite{wang2014exploring} are in line with those reported in the reference papers, and are mainly due to the weak performance achieved by these methods in the scenario $QF_1 > QF_2$ (while the performance when $QF_1 < QF_2$ are good, in their operative scenarios).

\begin{table}
\renewcommand\arraystretch{1.2}
\centering
\caption{The detection accuracy of the proposed method. }
	\vspace{0.1cm}
{\begin{tabular}{|p{1.5cm}<{\centering}|p{1.5cm}<{\centering}|p{1.5cm}<{\centering}|}
            \hline
            $k \setminus \hat{k}_r$ & $\hat{k}_r=1$ & $\hat{k}_r>1$ \\ \hline
			$k=1$ & {\bf 0.95} & 0.05  \\ \hline
            $k>1$ & 0.03 & {\bf 0.97}  \\ \hline

	\end{tabular}}
\label{tab.Accuracy}
\end{table}

\subsection{Localization performance ($k > 1$)} 



The results we obtained for tampering localization, averaged on the TP images only, that is the images correctly detected as tampered, are reported in Table \ref{tab.AvgLocPerformance}.
Not surprisingly, the proposed method works better than method \cite{bianchi2012image} in the Type II scenario, since the CNN-based $Q_1$ estimator is designed to work particularly for the NA-DJPEG case (the aligned case is assumed to occur with probability 1/64). The performances in the Type I scenario are also good and slightly better than those achieved by the best performing method \cite{wang2014exploring} on the average.
\begin{table}

\centering
\renewcommand\tabcolsep{3pt}
\caption{Comparison of the detection performance of our method with \cite{bianchi2012image} and \cite{wang2014exploring}. The table reports the TPR when FPR=0.05 over $\mathcal{D}_{ts}$, $\mathcal{D}_{ts}$ (Type I only) and $\mathcal{D}_{ts}$ (Type II only)}
	\vspace{0.1cm}
	{\begin{tabular}{|c|c|c|c|c|c|c|c|}
            \hline
               & $\mathcal{D}_{ts}$ & \multicolumn{3}{c|}{$\mathcal{D}_{ts}$ (Type I)} & \multicolumn{3}{c|}{$\mathcal{D}_{ts}$ (Type II)} \\ \hline
                 & & {\scriptsize All}  & {\tiny QF$_1$ $<$ QF$_2$} &  {\tiny QF$_1$ $>$ QF$_2$} &  {\scriptsize All}  &  {\tiny QF$_1$ $<$ QF$_2$} &  {\tiny QF$_1$ $>$ QF$_2$} \\ \hline
			\cite{bianchi2012image}-Al  & --& 0.31 &0.50& 0.14  & \em{--} & \em{--} & \em{--} \\ \hline
			\cite{bianchi2012image}-NAl   & -- & \em{--} & \em{--} & \em{--} & 0.18 & 0.25 & 0.11 \\ \hline
            \cite{wang2014exploring}   & --  & 0.39& 0.61 & 0.17 & -- & -- & --   \\ \hline
            Our   & 0.97 & \textbf{0.97}& \textbf{0.97} & \textbf{0.96} &\textbf{0.96} & \textbf{0.97} & \textbf{0.95} \\ \hline
	\end{tabular}}
	\label{tab.DetPerformance}
	\vspace{-0.1cm}
\end{table}

\begin{table}
\centering
\renewcommand\tabcolsep{3pt}
\caption{Average localization performance of the methods (MCC), averaged on the set of TP images for each method.}
	\vspace{0.1cm}
	{\begin{tabular}{|c|c|c|c|c|c|c|c|}
            \hline
               & $\mathcal{D}_{ts}$ & \multicolumn{3}{c|}{$\mathcal{D}_{ts}$ (Type I)} & \multicolumn{3}{c|}{$\mathcal{D}_{ts}$ (Type II)} \\ \hline
                 & & {\scriptsize All}  & {\tiny QF$_1$ $<$ QF$_2$} &  {\tiny QF$_1$ $>$ QF$_2$} &  {\scriptsize All}  &  {\tiny QF$_1$ $<$ QF$_2$} &  {\tiny QF$_1$ $>$ QF$_2$} \\ \hline
			\cite{bianchi2012image}-Al  & -- & 0.63 &0.77& 0.09  & \em{--} & \em{--} & \em{--} \\ \hline
			\cite{bianchi2012image}-NAl   & --& \em{--} & \em{--} & \em{--} & 0.47 & 0.58& 0.20 \\ \hline
            \cite{wang2014exploring}   & --&0.64 & \textbf{0.81}& 0.02 & -- & -- & --  \\ \hline
             Our   & 0.64 & \textbf{0.65}& 0.61 & \textbf{0.69} &\textbf{0.63} & \textbf{0.59} & \textbf{0.67} \\ \hline
	\end{tabular}}
	\label{tab.AvgLocPerformance}
\end{table}
The capability to work both in the case of A-DJPEG and NA-DJPEG is a noticeable property of our method, since the information about the alignment of the compression grids is not in general available.

Tables \ref{tab.LocPerformancek2} through \ref{tab.LocPerformancek4}
show the localization results of the proposed method for the NA-DJPEG case in the various settings, for various combinations of the quality factors of the background and foreground regions.
Specifically, Table \ref{tab.LocPerformancek2} reports the results for $k = 2$, for two different tampering sizes, namely $96 \times 96$ and $128 \times 128$.  Table \ref{tab.LocPerformancek3} and Table \ref{tab.LocPerformancek4} report the results for $k = 3$ and $k = 4$, for two different values of the quality factor of the background $QF_1$ of the background, i.e., $QF_1 = 85$ and $95$,  when the tampering size is equal to $128 \times 128$.
Similarly, Tables \ref{tab.LocPerformancek2_AL} through \ref{tab.LocPerformancek4_AL}
show the localization results for the A-DJPEG case, for  $k=2,3$ and $4$ respectively.
The sets  $\mathcal{D}_I$ and $\mathcal{D}_{II}$ are considered for these tests, respectively in the NA-DJPEG and A-DJPEG case.

Expectedly, better results are achieved when the tampering size is large (see Tables \ref{tab.LocPerformancek2} and \ref{tab.LocPerformancek3}), for $k=2$. The average difference in the MCC between the tampering size of $96 \times 96$ and $128 \times 128$ for the other values of $k$ is similar, ranging from $0.030$ to $0.090$ depending on the setting.
%
We can observe that our method greatly outperforms \cite{bianchi2012image} in all the settings for the NA-DJPEG case.
Regarding the performance in the A-DJPEG scenario, the method in \cite{wang2014exploring} always outperforms \cite{bianchi2012image} (the A-DJPEG method) when $k> 2$ and $QF_1 < 90$ ($QF_2$),  while
when $QF_1 > QF_2$ both methods can not correctly localize tampering.
Compared to our method, the performance of \cite{wang2014exploring} are superior in all the cases when the background $QF_1$ is smaller than $90$, with a gain in the MCC which is about 0.1\footnote{These values are not directly comparable since they are averaged on a different image sets, which in the case of the proposed method is much larger.}.
The performance loss in these cases is the price to pay for a general method, that can work in all the settings of $QF_{1,i}$ and $QF_2$,  and focuses on the more probable NA-DJPEG scenario (and then is not specifically designed for the A-DJPEG scenario).



\begin{table*}[htbp]
\renewcommand\arraystretch{1.3}
\scriptsize
\caption{Localization performance (MCC) for $k=2$, for the cases of tampering size $96\times 96$ (left) and $128\times 128$ (right). Performance are measured on  $\mathcal{D}_{II}$. The number of TP images is reported in brackets.
}
\centering
\subtable{
\begin{tabular}
  {|p{0.15cm}<{\centering}|p{0.8cm}<{\centering}|p{0.9cm}<{\centering}|p{0.9cm}<{\centering}|p{0.9cm}<{\centering}
 |p{0.9cm}<{\centering}|p{0.9cm}<{\centering}|}
  \hline
  \multicolumn{2}{|c|}{\diagbox[width=1.5cm]{$QF_{1}$}{$QF_{1,2}$}} &65& 75 &85  & 95& 98 \\
  \hline  \hline
  \multirow{2}{*}{75}&\text{\cite{bianchi2012image}-NAl}&\textbf{0.659}(10)&---&0.643(9)&0.650(13)&0.667(12)\\\cline{2-7}
  &Our&0.428(92)&---&\textbf{0.701}(99)& \textbf{0.794}(99)& \textbf{0.799}(100)\\ \hline \hline
  \multirow{2}{*}{85}&\text{\cite{bianchi2012image}-NAl}&0.102(3)&0.091(14)&---&0.114(14)&0.123(10)\\\cline{2-7}
  &Our&\textbf{0.738}(98)&\textbf{0.558}(98)&---& \textbf{0.646}(94)& \textbf{0.708}(97)\\ \hline \hline
  \multirow{2}{*}{95}&\text{\cite{bianchi2012image}-NAl}&0.086(12)&0.103(15)&0.108(17)&---&0.153(13)\\\cline{2-7}
  &Our&\textbf{0.805}(98)&\textbf{0.769}(98)& \textbf{0.665}(95)&---& \textbf{0.159}(19)\\ \hline \hline
  \multirow{2}{*}{98}&\text{\cite{bianchi2012image}-NAl}&0.090(17)&0.151(18)&0.140(8)&0.102(13)&---\\\cline{2-7}
  &Our&\textbf{0.804}(96)&\textbf{0.777}(98)& \textbf{0.698}(94)& \textbf{0.421}(41)&---\\ \hline
  \end{tabular}%

}
\subtable{
 \begin{tabular}
  {|p{0.15cm}<{\centering}|p{0.8cm}<{\centering}|p{0.9cm}<{\centering}|p{0.9cm}<{\centering}|p{0.9cm}<{\centering}
 |p{0.9cm}<{\centering}|p{0.9cm}<{\centering}|}
  \hline
  \multicolumn{2}{|c|}{\diagbox[width=1.5cm]{$QF_{1}$}{$QF_{1,2}$}} &65& 75 &85  & 95& 98 \\
  \hline  \hline
  \multirow{2}{*}{75}&\text{\cite{bianchi2012image}-NAl}&\textbf{0.775}(25)&---&0.744(30)&0.540(26)&0.741(31)\\\cline{2-7}
  &Our&0.542(97)&---&\textbf{0.771}(100)& \textbf{0.833}(100)& \textbf{0.845}(100)\\ \hline \hline
  \multirow{2}{*}{85}&\text{\cite{bianchi2012image}-NAl}&0.179(10)&0.201(9)&---&0.230(9)&0.286(8)\\\cline{2-7}
  &Our&\textbf{0.786}(100)&\textbf{0.701}(98)&---& \textbf{0.694}(99)& \textbf{0.757}(100)\\ \hline \hline
  \multirow{2}{*}{95}&\text{\cite{bianchi2012image}-NAl}&0.186(16)&0.183(12)&0.199(9)&---&\textbf{0.240}(8)\\\cline{2-7}
  &Our&\textbf{0.826}(99)&\textbf{0.847}(100)& \textbf{0.703}(98)&---&0.136(27)\\ \hline \hline
  \multirow{2}{*}{98}&\text{\cite{bianchi2012image}-NAl}&0.223(14)&0.131(5)&0.103(7)&0.175(8)&---\\\cline{2-7}
  &Our&\textbf{0.835}(100)&\textbf{0.850}(100)& \textbf{0.778}(100)& \textbf{0.546}(100)&---\\ \hline
  \end{tabular}%

}
\label{tab.LocPerformancek2}
\end{table*}

\begin{table*}[htbp]
\renewcommand\arraystretch{1.3}
\scriptsize
\caption{Localization performance (MCC) for  $k=3$, $128\times 128$,  $QF_{1} = 85$ (left), $QF_{1} = 95$ (right). Performance are measured on  $\mathcal{D}_{II}$. The number of TP images is reported in brackets.}
\centering
\subtable{
  \begin{tabular}
    {|p{0.18cm}<{\centering}|p{0.8cm}<{\centering}|p{1.3cm}<{\centering}|p{1.3cm}<{\centering}|p{1.3cm}<{\centering}
  |p{1.3cm}<{\centering}|}
  \hline
  \multicolumn{2}{|c|}{\diagbox[width=1.5cm]{$QF_{1,2}$}{$QF_{1,3}$}} &65& 75 &95  & 98 \\
  \hline  \hline
  \multirow{2}{*}{65}&\text{\cite{bianchi2012image}-NAl}&---&0.198(11)&0.356(7)&0.232(12)\\\cline{2-6}
  &Our&---&\textbf{0.728}(100)&\textbf{0.634}(100) &\textbf{0.684}(100)\\ \hline\hline
  \multirow{2}{*}{75}&\text{\cite{bianchi2012image}-NAl}&0.127(10)&---&0.254(12)&0.243(9)\\\cline{2-6}
  &Our&\textbf{0.717}(100)&--- &\textbf{0.614}(99)&\textbf{0.636}(100)\\ \hline\hline
  \multirow{2}{*}{95}&\text{\cite{bianchi2012image}-NAl}&0.221(12)&0.192(18)&---&0.356(11)\\\cline{2-6}
  &Our&\textbf{0.667}(100)&\textbf{0.621}(100)&--- &\textbf{0.701}(99)\\ \hline\hline
  \multirow{2}{*}{98}&\text{\cite{bianchi2012image}-NAl}&0.376(8)&0.176(16)&0.329(8)&---\\\cline{2-6}
  &Our&\textbf{0.705}(100)&\textbf{0.626}(100) &\textbf{0.694}(99)&---\\ \hline
  \end{tabular}%

}
\subtable{
  \begin{tabular}
    {|p{0.18cm}<{\centering}|p{0.8cm}<{\centering}|p{1.3cm}<{\centering}|p{1.3cm}<{\centering}|p{1.3cm}<{\centering}
  |p{1.3cm}<{\centering}|}
  \hline
  \multicolumn{2}{|c|}{\diagbox[width=1.5cm]{$QF_{1,2}$}{$QF_{1,3}$}} &65& 75 &85  & 98 \\
  \hline  \hline
  \multirow{2}{*}{65}&\text{\cite{bianchi2012image}-NAl}&---&0.168(15)&0.178(15)&0.260(23)\\\cline{2-6}
  &Our&---&\textbf{0.814}(100)&\textbf{0.679}(100) &\textbf{0.565}(100)\\ \hline\hline
  \multirow{2}{*}{75}&\text{\cite{bianchi2012image}-NAl}&0.292(9)&---&0.148(11)&0.255(22)\\\cline{2-6}
  &Our&\textbf{0.817}(100)&--- &\textbf{0.720}(100)&\textbf{0.560}(98)\\ \hline\hline
  \multirow{2}{*}{85}&\text{\cite{bianchi2012image}-NAl}&0.198(18)&0.336(17)&---&0.137(11)\\\cline{2-6}
  &Our&\textbf{0.688}(100)&\textbf{0.694}(99)&--- &\textbf{0.463}(98)\\ \hline\hline
  \multirow{2}{*}{98}&\text{\cite{bianchi2012image}-NAl}&0.145(11)&0.148(14)&0.274(13)&---\\\cline{2-6}
  &Our&\textbf{0.595}(100)&\textbf{0.584}(100) &\textbf{0.498}(100)&---\\ \hline
  \end{tabular}%

}
\label{tab.LocPerformancek3}
\vspace{-0.2cm}
\end{table*}

\begin{table*}[htbp]
\renewcommand\arraystretch{1.3}
\scriptsize
\vspace{-0.2cm}
\caption{Localization performance (MCC) for $k=4$, $128\times 128$, $QF_{1} = 85$ (left), $QF_{1} = 95$ (right).  $QF_{1,2}$ is set to $60$. Performance are measured on  $\mathcal{D}_{II}$. The number of TP images is reported in brackets.}
\centering
\subtable{
  \begin{tabular}
    {|p{0.18cm}<{\centering}|p{0.8cm}<{\centering}|p{1.3cm}<{\centering}|p{1.3cm}<{\centering}|p{1.3cm}<{\centering}
  |p{1.3cm}<{\centering}|}
  \hline
  \multicolumn{2}{|c|}{\diagbox[width=1.5cm]{$QF_{1,3}$}{$QF_{1,4}$}} &65& 75 &95  & 98 \\
  \hline  \hline
  \multirow{2}{*}{65}&\text{\cite{bianchi2012image}-NAl}&---&0.322(14)&0.240(14)&0.320(21)\\\cline{2-6}
  &Our&---&\textbf{0.741}(100)&\textbf{0.707}(100) &\textbf{0.710}(100)\\ \hline\hline
  \multirow{2}{*}{75}&\text{\cite{bianchi2012image}-NAl}&0.381(14)&---&0.309(16)&0.410(12)\\\cline{2-6}
  &Our&\textbf{0.709}(100)&--- &\textbf{0.664}(100)&\textbf{0.659}(100)\\ \hline\hline
  \multirow{2}{*}{95}&\text{\cite{bianchi2012image}-NAl}&0.334(11)&0.293(17)&---&0.335(17)\\\cline{2-6}
  &Our&\textbf{0.689}(100)&\textbf{0.656}(100)&--- &\textbf{0.724}(100)\\ \hline\hline
  \multirow{2}{*}{98}&\text{\cite{bianchi2012image}-NAl}&0.288(17)&0.235(13)&0.269(11)&---\\\cline{2-6}
  &Our&\textbf{0.757}(100)&\textbf{0.692}(100) &\textbf{0.698}(100)&---\\ \hline
  \end{tabular}%

}
\subtable{
\begin{tabular}
    {|p{0.18cm}<{\centering}|p{0.8cm}<{\centering}|p{1.3cm}<{\centering}|p{1.3cm}<{\centering}|p{1.3cm}<{\centering}
  |p{1.3cm}<{\centering}|}
  \hline
  \multicolumn{2}{|c|}{\diagbox[width=1.5cm]{$QF_{1,3}$}{$QF_{1,4}$}} &65& 75 &85  & 98 \\
  \hline  \hline
  \multirow{2}{*}{65}&\text{\cite{bianchi2012image}-NAl}&---&0.198(20)&0.228(19)&0.297(13)\\\cline{2-6}
  &Our&---&\textbf{0.807}(100)&\textbf{0.738}(100) &\textbf{0.646}(100)\\ \hline\hline
  \multirow{2}{*}{75}&\text{\cite{bianchi2012image}-NAl}&0.211(18)&---&0.371(15)&0.277(13)\\\cline{2-6}
  &Our&\textbf{0.797}(100)&--- &\textbf{0.734}(100)&\textbf{0.625}(100)\\ \hline\hline
  \multirow{2}{*}{85}&\text{\cite{bianchi2012image}-NAl}&0.280(16)&0.290(15)&---&0.233(25)\\\cline{2-6}
  &Our&\textbf{0.730}(100)&\textbf{0.719}(100)&--- &\textbf{0.477}(100)\\ \hline\hline
  \multirow{2}{*}{98}&\text{\cite{bianchi2012image}-NAl}&0.239(17)&0.140(17)&0.266(19)&---\\\cline{2-6}
  &Our&\textbf{0.643}(100)&\textbf{0.632}(100) &\textbf{0.494}(100)&---\\ \hline
  \end{tabular}%

}
\label{tab.LocPerformancek4}
\vspace{-0.2cm}
\end{table*}

\begin{table*}[htbp]
\renewcommand\arraystretch{1.3}
\scriptsize
\caption{Localization performance (MCC) for $k=2$,  $96\times 96$ (left) and $128\times 128$ (right). Performance are measured on  $\mathcal{D}_{I}$. The number of TP images is reported in brackets.}
\centering
\subtable{
\begin{tabular}
    {|p{0.15cm}<{\centering}|p{0.8cm}<{\centering}|p{0.9cm}<{\centering}|p{0.9cm}<{\centering}|p{0.9cm}<{\centering}
 |p{0.9cm}<{\centering}|p{0.9cm}<{\centering}|}
  \hline
  \multicolumn{2}{|c|}{\diagbox[width=1.5cm]{$QF_{1}$}{$QF_{1,2}$}} &65& 75 &85  & 95& 98 \\
  \hline  \hline
  \multirow{3}{*}{75}&\text{\cite{bianchi2012image}-Al}&0.878(10)&---&0.862(9)&0.851(13)& \textbf{0.888}(12)\\\cline{2-7}
  &\text{\cite{wang2014exploring}}&\textbf{0.891}(60)&---&\textbf{0.885}(54) &\textbf{0.875}(70)& 0.879(68)\\ \cline{2-7}
  &Our&0.476(98)&---&0.744(100)& 0.796(100)& 0.814(99)\\ \hline \hline
  \multirow{3}{*}{85}&\text{\cite{bianchi2012image}-Al}&0.417(11)&0.608(13)&---&0.463(19)&0.590(22)\\\cline{2-7}
  &\text{\cite{wang2014exploring}}&\textbf{0.725}(40)&\textbf{0.848}(50)&---&0.706(47)& \textbf{0.781}(43)\\\cline{2-7}
  &Our&0.710(100)&0.523(93)&---& \textbf{0.738}(98)&0.734(100)\\ \hline \hline
  \multirow{3}{*}{95}&\text{\cite{bianchi2012image}-Al}&0.044(6)&0.054(14)&0.069(21)&---&0.081(15)\\\cline{2-7}
  &\text{\cite{wang2014exploring}}&0.002(24)&0.051(14) &0.056(11)&---&0.000(16)\\ \cline{2-7}
  &Our&\textbf{0.792}(99)&\textbf{0.775}(100)& \textbf{0.717}(98)&---& \textbf{0.210}(34)\\ \hline \hline
  \multirow{3}{*}{98}&\text{\cite{bianchi2012image}-Al}&0.039(15)&0.023(6)&0.163(12)&0.027(13)&---\\\cline{2-7}
  &\text{\cite{wang2014exploring}}&0.000(18)&0.000(10) &0.000(14)&0.006(21)&---\\ \cline{2-7}
  &Our&\textbf{0.804}(100)&\textbf{0.784}(99)& \textbf{0.680}(94)& \textbf{0.377}(34)&---\\ \hline
  \end{tabular}%

}
\subtable{
 \begin{tabular}
    {|p{0.15cm}<{\centering}|p{0.8cm}<{\centering}|p{0.9cm}<{\centering}|p{0.9cm}<{\centering}|p{0.9cm}<{\centering}
 |p{0.9cm}<{\centering}|p{0.9cm}<{\centering}|}
  \hline
  \multicolumn{2}{|c|}{\diagbox[width=1.5cm]{$QF_{1}$}{$QF_{1,2}$}} &65& 75 &85  & 95& 98 \\
  \hline  \hline
  \multirow{3}{*}{75}&\text{\cite{bianchi2012image}-Al}&0.864(58)&---&0.872(74)&0.856(66)&0.857(65)\\\cline{2-7}
  &\text{\cite{wang2014exploring}}&\textbf{0.893}(58)&---&\textbf{0.891}(71) &\textbf{0.892}(71)&\textbf{0.893}(73)\\ \cline{2-7}
  &Our&0.581(99)&---&0.808(100)& 0.846(100)& 0.853(100)\\ \hline \hline
  \multirow{3}{*}{85}&\text{\cite{bianchi2012image}-Al}&0.618(41)&0.606(34)&---&0.621(34)&0.617(42)\\\cline{2-7}
  &\text{\cite{wang2014exploring}}&0.732(57)&\textbf{0.791}(52)&---&\textbf{0.823}(53)& \textbf{0.797}(64)\\\cline{2-7}
  &Our&\textbf{0.810}(100)&0.650(100)&---&0.821(98)&0.797(99)\\ \hline \hline
  \multirow{3}{*}{95}&\text{\cite{bianchi2012image}-Al}&0.049(15)&0.075(11)&0.096(9)&---&0.059(16)\\\cline{2-7}
  &\text{\cite{wang2014exploring}}&0.000(100)&0.011(18) &0.020(11)&---&0.039(21)\\ \cline{2-7}
  &Our&\textbf{0.850}(100)&\textbf{0.833}(100)& \textbf{0.702}(99)&---& \textbf{0.137}(36)\\ \hline \hline
  \multirow{3}{*}{98}&\text{\cite{bianchi2012image}-Al}&0.099(12)&0.043(5)&0.188(15)&0.101(10)&---\\\cline{2-7}
  &\text{\cite{wang2014exploring}}&0.051(15)&0.012(13) &0.007(20)&0.006(19)&---\\ \cline{2-7}
  &Our&\textbf{0.857}(100)&\textbf{0.840}(100)& \textbf{0.772}(99)& \textbf{0.425}(49)&---\\ \hline
  \end{tabular}%

}
\label{tab.LocPerformancek2_AL}
\vspace{-0.2cm}
\end{table*}

\begin{table*}[htbp]
\vspace{-0.2cm}
\renewcommand\arraystretch{1.3}
\scriptsize
\caption{Localization performance (MCC) for $k=3$, $128\times 128$, $QF_1=85$ (left), $QF_1=95$ (right). Performance are measured on  $\mathcal{D}_{I}$. The number of TP images is reported in brackets.}
\centering
\subtable{
 \begin{tabular}
    {|p{0.18cm}<{\centering}|p{0.8cm}<{\centering}|p{1.3cm}<{\centering}|p{1.3cm}<{\centering}|p{1.3cm}<{\centering}
  |p{1.3cm}<{\centering}|}
  \hline
  \multicolumn{2}{|c|}{\diagbox[width=1.5cm]{$QF_{1,2}$}{$QF_{1,3}$}} &65& 75 & 95& 98 \\
  \hline  \hline
  \multirow{3}{*}{65}&\text{\cite{bianchi2012image}-Al}&---&0.641(43)&0.695(45)&0.722(51)\\\cline{2-6}
  &\text{\cite{wang2014exploring}}&---&\textbf{0.782}(50)&\textbf{0.768}(68) &\textbf{0.829}(57)\\ \cline{2-6}
  &Our&---&0.685(100)&0.671(100) &0.686(100)\\ \hline\hline
  \multirow{3}{*}{75}&\text{\cite{bianchi2012image}-Al}&0.725(45)&---&0.771(48)&0.676(51)\\\cline{2-6}
  &\text{\cite{wang2014exploring}}&\textbf{0.781}(57)&---&\textbf{0.842}(60) &\textbf{0.732}(64)\\ \cline{2-6}
  &Our&0.671(100)&---&0.654(99)& 0.636(100)\\ \hline \hline
  \multirow{3}{*}{95}&\text{\cite{bianchi2012image}-Al}&0.714(47)&0.602(46)&---&0.698(41)\\\cline{2-6}
  &\text{\cite{wang2014exploring}}&\textbf{0.816}(58)&\textbf{0.798}(58) &---&\textbf{0.838}(65)\\ \cline{2-6}
  &Our&0.683(100)&0.618(100)&--- &0.768(100)\\ \hline \hline
  \multirow{3}{*}{98}&\text{\cite{bianchi2012image}-Al}&0.706(42)&0.623(43)&0.687(49)&---\\\cline{2-6}
  &\text{\cite{wang2014exploring}}&\textbf{0.771}(60)&\textbf{0.740}(59) &0.791(54)&---\\ \cline{2-6}
  &Our&0.656(100)&0.616(99)&\textbf{ 0.798}(94)&---\\ \hline
  \end{tabular}%

}
\subtable{
 \begin{tabular}
    {|p{0.18cm}<{\centering}|p{0.8cm}<{\centering}|p{1.3cm}<{\centering}|p{1.3cm}<{\centering}|p{1.3cm}<{\centering}
  |p{1.3cm}<{\centering}|}
  \hline
  \multicolumn{2}{|c|}{\diagbox[width=1.5cm]{$QF_{1,2}$}{$QF_{1,3}$}} &65& 75 & 85& 98 \\
  \hline  \hline
  \multirow{3}{*}{65}&\text{\cite{bianchi2012image}-Al}&---&0.133(15)&0.028(14)&0.116(19)\\\cline{2-6}
  &\text{\cite{wang2014exploring}}&---&0.000(9)&0.034(17) &0.063(19)\\ \cline{2-6}
  &Our&---&\textbf{0.799}(100)&\textbf{0.726}(100) &\textbf{0.564}(100)\\ \hline\hline
  \multirow{3}{*}{75}&\text{\cite{bianchi2012image}-Al}&0.136(12)&---&0.089(19)&0.056(22)\\\cline{2-6}
  &\text{\cite{wang2014exploring}}&0.000(12)&---&0.044(16) &0.000(14)\\ \cline{2-6}
  &Our&\textbf{0.802}(100)&---&\textbf{0.758}(100)& \textbf{0.550}(98)\\ \hline \hline
  \multirow{3}{*}{85}&\text{\cite{bianchi2012image}-Al}&0.092(15)&0.128(21)&---&0.097(11)\\\cline{2-6}
  &\text{\cite{wang2014exploring}}&0.059(11)&0.030(25) &---&0.078(14)\\ \cline{2-6}
  &Our&\textbf{0.697}(99)&\textbf{0.749}(100)&--- &\textbf{0.539}(97)\\ \hline \hline
  \multirow{3}{*}{98}&\text{\cite{bianchi2012image}-Al}&0.090(12)&0.110(12)&0.083(15)&---\\\cline{2-6}
  &\text{\cite{wang2014exploring}}&0.000(12)&0.035(18) &0.026(18)&---\\ \cline{2-6}
  &Our&\textbf{0.593}(100)&\textbf{0.578}(100)& \textbf{0.545}(100)&---\\ \hline
  \end{tabular}%

}
\label{tab.LocPerformancek3_AL}
\end{table*}

\begin{table*}[htbp]
\vspace{-0.2cm}
\renewcommand\arraystretch{1.3}
\scriptsize
\caption{Localization performance (MCC) for $k=4$, $128\times 128$, $QF_1=85$ (left) and $QF_1=95$ (right), $QF_{1,2}$ is set to $60$. Performance are measured on  $\mathcal{D}_{I}$. The number of TP images is reported in brackets.}
\centering
\subtable{
 \begin{tabular}
    {|p{0.18cm}<{\centering}|p{0.8cm}<{\centering}|p{1.3cm}<{\centering}|p{1.3cm}<{\centering}|p{1.3cm}<{\centering}
  |p{1.3cm}<{\centering}|}
  \hline
  \multicolumn{2}{|c|}{\diagbox[width=1.5cm]{$QF_{1,3}$}{$QF_{1,4}$}} &65& 75 & 95& 98 \\
  \hline  \hline
  \multirow{3}{*}{65}&\text{\cite{bianchi2012image}-Al}&---&0.679(40)&0.619(41)&0.592(40)\\\cline{2-6}
  &\text{\cite{wang2014exploring}}&---&\textbf{0.802}(53)&\textbf{0.762}(54) &\textbf{0.767}(51)\\ \cline{2-6}
  &Our&---&0.695(100)&0.707(100) &0.714(100)\\ \hline\hline
  \multirow{3}{*}{75}&\text{\cite{bianchi2012image}-Al}&0.646(50)&---&0.639(44)&0.648(47)\\\cline{2-6}
  &\text{\cite{wang2014exploring}}&\textbf{0.792}(58)&---&\textbf{0.780}(50) &\textbf{0.787}(55)\\ \cline{2-6}
  &Our&0.709(100)&---&0.664(100)& 0.679(100)\\ \hline \hline
  \multirow{3}{*}{95}&\text{\cite{bianchi2012image}-Al}&0.682(39)&0.672(39)&---&0.710(39)\\\cline{2-6}
  &\text{\cite{wang2014exploring}}&\textbf{0.766}(52)&\textbf{0.802}(58) &---&\textbf{0.793}(51)\\ \cline{2-6}
  &Our&0.702(100)&0.647(100)&--- &0.720(100)\\ \hline \hline
  \multirow{3}{*}{98}&\text{\cite{bianchi2012image}-Al}&0.673(38)&0.684(43)&0.687(41)&---\\\cline{2-6}
  &\text{\cite{wang2014exploring}}&\textbf{0.769}(55)&\textbf{0.796}(44) &\textbf{0.814}(57)&---\\ \cline{2-6}
  &Our&0.724(100)&0.693(100)& 0.725(100)&---\\ \hline
  \end{tabular}%

}
\subtable{
 \begin{tabular}
  {|p{0.18cm}<{\centering}|p{0.8cm}<{\centering}|p{1.3cm}<{\centering}|p{1.3cm}<{\centering}|p{1.3cm}<{\centering}
  |p{1.3cm}<{\centering}|}
  \hline
  \multicolumn{2}{|c|}{\diagbox[width=1.5cm]{$QF_{1,3}$}{$QF_{1,4}$}} &65& 75 & 85& 98 \\
  \hline  \hline
  \multirow{3}{*}{65}&\text{\cite{bianchi2012image}-Al}&---&0.135(22)&0.170(16)&0.221(14)\\\cline{2-6}
  &\text{\cite{wang2014exploring}}&---&0.000(8)&0.152(8) &0.033(12)\\ \cline{2-6}
  &Our&---&\textbf{0.803}(100)&\textbf{0.755}(100) &\textbf{0.625}(100)\\ \hline\hline
  \multirow{3}{*}{75}&\text{\cite{bianchi2012image}-Al}&0.119(12)&---&0.054(17)&0.118(8)\\\cline{2-6}
  &\text{\cite{wang2014exploring}}&0.091(13)&---&0.081(8) &0.000(15)\\ \cline{2-6}
  &Our&\textbf{0.821}(100)&---&\textbf{0.770}(100)& \textbf{0.621}(100)\\ \hline \hline
  \multirow{3}{*}{85}&\text{\cite{bianchi2012image}-Al}&0.113(11)&0.065(21)&---&0.142(20)\\\cline{2-6}
  &\text{\cite{wang2014exploring}}&0.090(9)&0.088(17) &---&0.054(9)\\ \cline{2-6}
  &Our&\textbf{0.774}(100)&\textbf{0.752}(100)&--- &\textbf{0.483}(100)\\ \hline \hline
  \multirow{3}{*}{98}&\text{\cite{bianchi2012image}-Al}&0.105(22)&0.113(8)&0.069(16)&---\\\cline{2-6}
  &\text{\cite{wang2014exploring}}&0.084(11)&0.054(13) &0.058(14)&---\\ \cline{2-6}
  &Our&\textbf{0.637}(100)&\textbf{0.668}(100)& \textbf{0.534}(100)&---\\ \hline
  \end{tabular}%

}
\label{tab.LocPerformancek4_AL}
\vspace{-0.1cm}
\end{table*}

\begin{figure}[!htbp]
\centering\includegraphics[width=0.9\columnwidth]{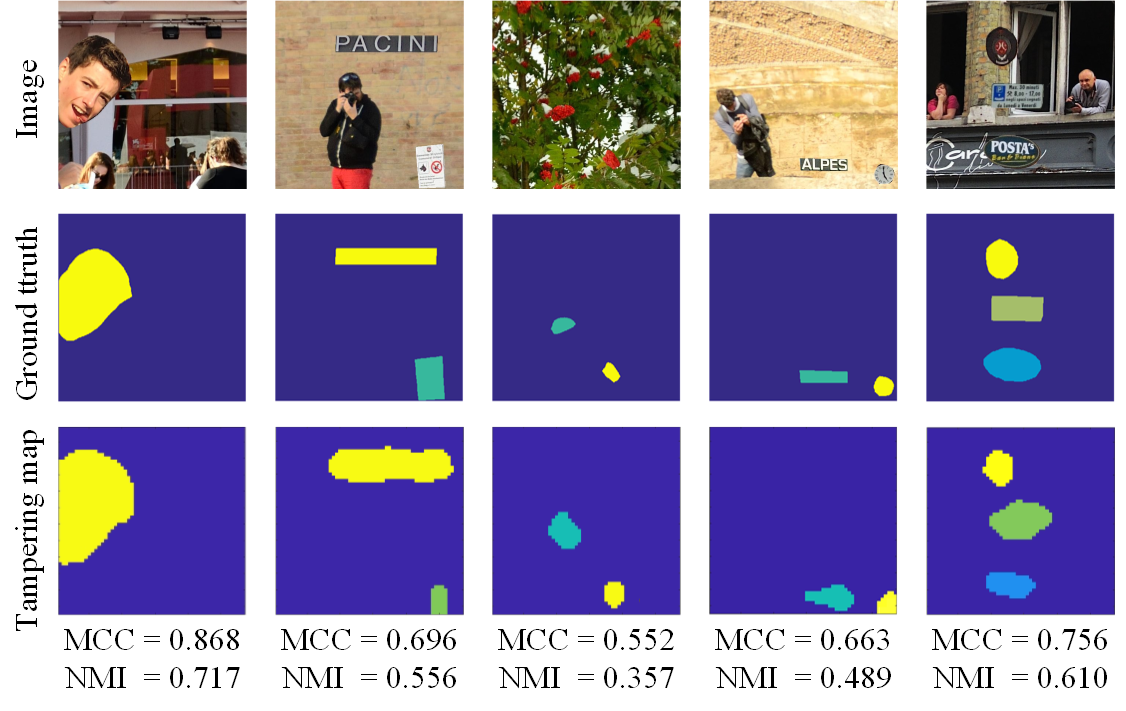}
\caption{Examples of the results provided by our system on some tampered images, for the case of A-DJPEG, with $QF_1 < QF_2$. The ground truth and the output maps produced by our method are reported, along with the MCC and NMI values.
}
\label{fig_1}
\end{figure}

\subsection{Attribution performance ($k > 1$)}
\label{sec.attribution}

As opposed to state-of-the-art methods, the system introduced in this paper allows to distinguish spliced regions coming from different donor images.
Fig. \ref{fig_1} shows some examples of the output maps obtained on some tampered images with different $k$. The corresponding NMI value measuring the goodness of the clustering (see Section \ref{sec.eval_metrics}) is also reported.
We can see that, even if the NMI indexes can theoretically reach 1, satisfactory clustering results are already obtained with much lower NMI values.

The average NMI values obtained over the test set $\mathcal{D}_{ts}$ and the subsets of Type I and Type II images in $\mathcal{D}_{ts}$ are reported in Table \ref{tab.averageMNI}. The NMI values of the state-of-the-art methods are not reported since they do not distinguish spliced regions coming from different donor images.
%
The average NMI in the Type I and II cases are very similar. Notice that a value of the NMI around 0.5 is satisfactory, since the localization and clustering results are already good with an NMI around 0.6, while the NMI is close to zero when either the localization or the clustering result is poor, then the average NMI is not expected to be very high.
The clustering performance for some settings of $k> 2$, with  background JPEG quality $QF_1 = 85$ (measured on 100 TP images each) are reported in Table \ref{tab.CluPerformance}. Similar NMI values are obtained for the various combinations of quality factors of the foreground regions.
\begin{table}
\renewcommand\arraystretch{1.1}
\centering
\caption{Average attribution performance (NMI) of the proposed method. }
	\vspace{0.1cm}
	{\begin{tabular}{|p{1.5cm}<{\centering}|p{1.8cm}<{\centering}|p{1.8cm}<{\centering}|p{1.8cm}<{\centering}|}
            \hline
			Test set & $\mathcal{D}_{ts}$ &$\mathcal{D}_{ts}$ (Type I)& $\mathcal{D}_{ts}$ (Type II)\\ \hline
           NMI    &  0.475 &  0.481 & 0.468\\ \hline
	\end{tabular}}
\label{tab.averageMNI}
\end{table}

\begin{table*}[htbp]
\scriptsize
\caption{Attribution performance (NMI) of the proposed method for $k=3$ ((a) and (c)) and $k=4$ ((b) and (d)), tampering size $128\times 128$, $QF_1=85$. (a) and (b) refer to the non-aligned case (performance measured on  $\mathcal{D}_{II}$), (c) and (d) refer to the aligned case (performance measured on  $\mathcal{D}_{I}$). For $k = 4$, $QF_{1,2}$ is set to $60$.}
\centering
\setlength{\tabcolsep}{2.8pt}
\subtable[]{
 \begin{tabular}{|p{1cm}<{\centering}|c|c|c |c|}
  \hline
  \diagbox[width=1.2cm]{$QF_{1,2}$}{$QF_{1,3}$} &65& 75 & 95& 98 \\
  \hline
  65&---&0.559&0.497&0.532\\ \hline
  75&0.547&---&0.463& 0.482\\ \hline
  95&0.516&0.467&--- &0.488\\ \hline
  98&0.508&0.475& 0.476&---\\ \hline
  \end{tabular}%
       \label{tab:thirdtableNMI}
}
\subtable[]{
 \begin{tabular}{|p{1cm}<{\centering}|c|c|c |c|}
  \hline
 \diagbox[width=1.2cm]{$QF_{1,3}$}{$QF_{1,4}$} &65& 75 & 95& 98 \\
  \hline
  65&---&0.555&0.535 &0.536\\ \hline
  75&0.529&---&0.515& 0.503\\ \hline
  95&0.527&0.502&--- &0.521\\ \hline
  98&0.573&0.532& 0.503&---\\ \hline
  \end{tabular}%
       \label{tab:secondtable}
}
\subtable[]{\begin{tabular}{|p{1cm}<{\centering}|c|c|c |c|}
  \hline
  \diagbox[width=1.2cm]{$QF_{1,2}$}{$QF_{1,3}$} &65& 75 & 95& 98 \\
  \hline
  65&---&0.518&0.524 &0.531\\ \hline
  75&0.507&---&0.496& 0.682\\ \hline
  95&0.536&0.467&--- &0.553\\ \hline
  98&0.501&0.468& 0.553&---\\ \hline
  \end{tabular}%
 \label{tab:firsttableNMI}
 }
\subtable[]{
 \begin{tabular}{|p{1cm}<{\centering}|c|c|c |c|}
  \hline
 \diagbox[width=1.2cm]{$QF_{1,3}$}{$QF_{1,4}$} &65& 75 & 95& 98 \\
  \hline
  65&---&0.525&0.535 &0.548\\ \hline
  75&0.529&---&0.515& 0.523\\ \hline
  95&0.532&0.498&--- &0.514\\ \hline
  98&0.548&0.495& 0.523&---\\ \hline
  \end{tabular}%
       \label{tab:secondtableNMI}
}
\label{tab.CluPerformance}
\end{table*}


\subsection{Ablation study}
We also carried out an ablation study to measure the impact of the most important parameters of the system on localization and attribution accuracy. The results we got  on the dataset $\mathcal{D}_{ts}$ are reported in Table \ref{tab.ablation}. The performance of $K$-means clustering are also reported.

\begin{table}[h]
\centering
\renewcommand\tabcolsep{3pt}
\caption{Results of the ablation study (MCC/NMI).}
	\vspace{0.1cm}
	{\begin{tabular}{|p{2cm}<{\centering}|p{2cm}<{\centering}|p{2cm}<{\centering}|}
            \hline
			 & Type I & Type II \\\hline

            $\hat{k}$ + $K$-means & 0.567/0.409 & 0.520/0.374  \\\hline
            	        SC & 0.636/0.443 & 0.610/0.423  \\\hline
            $\hat{k}$ + SC & 0.643/0.470 & 0.622/0.456  \\\hline
            $\hat{k}$ + SC + MR& \textbf{0.652}/\textbf{0.481} & \textbf{0.633}/\textbf{0.468}  \\\hline
	\end{tabular}}
	\label{tab.ablation}
\end{table}
\begin{table}[h]
\centering
\renewcommand\tabcolsep{3pt}

\caption{Average localization performance (MCC) on the datasets in  \cite{bianchi2012image}  and \cite{wang2014exploring}.  TPR is reported within brackets.}
	\vspace{0.1cm}
  {\begin{tabular}{|p{1.5cm}<{\centering}|p{2cm}<{\centering}|p{2cm}<{\centering}|p{2cm}<{\centering}|}
            \hline
			Method & $\mathcal{D}_{{\scriptsize  \cite{bianchi2012image}},\text{A}}$ & $\mathcal{D}_{{\scriptsize \cite{bianchi2012image}},\text{NA}}$&$\mathcal{D}_{\scriptsize  \cite{wang2014exploring}}$\\\hline
            \cite{bianchi2012image}-Al & 0.657(0.46) &-- &0.664(0.46)\\\hline
	        \cite{bianchi2012image}-NAl &-- &0.518(0.53) &--\\\hline
            \cite{wang2014exploring} & 0.781(0.92) &-- &0.701(0.91)\\\hline
             Our & \textbf{0.904}(0.82) &\textbf{0.906}(0.80) &\textbf{0.712}(0.89)\\\hline
	\end{tabular}}
	\label{tab.stateofart}
	\vspace{-0.3cm}

\end{table}

The last line corresponds to  the proposed pipeline, including the step of refinement of the map (the final number of clusters is $\hat{k}_r$). We see that the best results can be obtained in this case.  Although the improvement in the NMI is not big,  it is a relevant one, given the poor sensitivity of this metric.
\vspace{-2mm}

\subsection{Results under mismatched testing conditions}

To test the generalization capability of our system, with particular reference to the CNN-based components, we run some experiments under different and more challenging testing conditions. The results of our tests are reported below.

\begin{table*}
\centering
\renewcommand\tabcolsep{3pt}
\caption{Average localization performance (MCC) on $\mathcal{D}_{Dr}$ for various tampering sizes, averaged on the set of TP images. The  TPR is reported within brackets.}
	\vspace{0.1cm}
%
%
 {\begin{tabular}{|p{1.2cm}<{\centering}|p{1.5cm}<{\centering}|p{1.5cm}<{\centering}|p{1.5cm}<{\centering}|
    p{1.5cm}<{\centering}|p{1.5cm}<{\centering}|p{1.5cm}<{\centering}|p{1.5cm}<{\centering}|p{1.5cm}<{\centering}|p{1.5cm}<{\centering}|}
            \hline
             \multirow{2}{*}{} & \multicolumn{3}{c|}{$96\times96$}& \multicolumn{3}{c|}{$128\times128$}& \multicolumn{3}{c|}{$156\times156$} \\
    \cline{2-10}
			 & All  & Type I&Tpye II& All  & Type I&Tpye II& All  & Type I&Tpye II\\\hline
           \cite{bianchi2012image}-Al & --- &0.532(0.46)& ---& --- &0.549(0.61)& ---& --- &0.570(0.63)& ---\\\hline
           \cite{bianchi2012image}-NAl & --- & ---&0.504(0.30)& --- & ---&0.507(0.26)& --- & ---&0.511(0.29)\\\hline
           \cite{wang2014exploring}& --- &0.544(0.50) &---& --- &0.573(0.39) &---& --- &0.589(0.61) &---\\\hline
           Our  & 0.568(0.97)&\textbf{0.579}(0.97) &\textbf{0.557}(0.97)& 0.641(0.98)&\textbf{0.645}(0.97) &\textbf{0.636}(0.98)& 0.683(0.98)&\textbf{0.693}(0.98) &\textbf{0.673}(0.97)\\\hline
	\end{tabular}}
	\label{tab.dresden}
\end{table*}

\begin{table}[t]
\renewcommand\arraystretch{1.1}
\centering
\caption{Average attribution performance (NMI) of the proposed method on  $\mathcal{D}_{Dr}$.}
	\vspace{0.1cm}
	{\begin{tabular}{|p{1.5cm}<{\centering}|p{1.8cm}<{\centering}|p{1.8cm}<{\centering}|p{1.8cm}<{\centering}|}
            \hline
			Test set & $\mathcal{D}_{Dr}$ &$\mathcal{D}_{Dr}$ (Type I)& $\mathcal{D}_{Dr}$ (Type II)\\ \hline
           NMI    &  0.481 &  0.486  &0.476 \\ \hline
	\end{tabular}}
\label{tab.averageMNI-dr}
\end{table}

\subsubsection{Mismatched datasets}

The results we got on the datasets used in \cite{bianchi2012image} and \cite{wang2014exploring}
are shown in Table \ref{tab.stateofart}.
By looking at this table, we see that the localization results of our method on the datasets $\mathcal{D}_{\footnotesize \cite{bianchi2012image}}$ (A and NA) and $\mathcal{D}_{\footnotesize \cite{wang2014exploring}}$ are very good, better than those achieved on $\mathcal{D}_{ts}$. The tamperings in these datasets are in fact performed under less challenging conditions compared to our home-made dataset, only in the case $k=2$, with a rather large tampering area and considering smaller  $QF_1$ values, facilitating the detection and localization tasks (in $\mathcal{D}_{ts}$, $QF_1$  is almost always above 75 and never goes below 60). It is also worth observing that our method is not designed for the DJPEG vs SJPEG scenario considered in these datasets\footnote{Notably, for single compressed regions, our estimator returns a quantization matrix of a very high compression quality ($QF > 98$). Therefore, when the method is applied for tampering localization in this scenario, the manipulation can still be exposed based on the inconsistencies between the estimated quantization steps of the background and the foreground.}.
%

The average localization results on the $\mathcal{D}_{Dr}$ datasets are reported in Table  \ref{tab.dresden}, for various tampering sizes. These results are similar to those achieved on $\mathcal{D}_{ts}$, confirming the good performance of our method even in the presence of dataset mismatch.
The localization performance for the various combinations of the quality factors of the background and foreground regions reveal that, similarly to what happen in the experiments with the matched dataset, the advantage of our method with respect to the state-of-the-art is stronger for large quality factors of the background. The results e are not reported for sake of brevity.
%

%
Attribution performance can also be measured on this dataset. The average NMI values obtained on $\mathcal{D}_{Dr}$
are reported in Table \ref{tab.averageMNI-dr} and are similar to those achieved on $\mathcal{D}_{ts}$.

\subsubsection{Non-standard quantization matrices}

%

To assess the behavior of the system with non-standard quantization matrices, we run some tests on images for which the first compression was carried out by using Photoshop (PS). The results  are reported in Table \ref{tab.PS}.

We see that our method can generalize to non-standard matrices and still outperform the methods in \cite{bianchi2012image} and \cite{wang2014exploring}. Table \ref{tab:ps:128} shows the results for the various combinations of PS qualities, for case 128 $\times$ 128. We see that the cases where the background has lower quality than the foreground, and they are close to each other, are the most difficult cases for our method. The detection performance are also poor in these cases.
Overall, we notice that these are
particularly relevant results since these test corresponds to a case of strong
mismatch between training and test data used for the CNN-based components and for the tuning of the parameters.
In principle, the results could be improved by training, or
fine-tuning, the models (and in particular, the CNN for $Q_1$ estimation) considering also non-standard quantization steps for the former compression.


\begin{table}
\centering
\renewcommand\tabcolsep{3pt}

\caption{Localization performance (MCC) when the first compression is done with Photoshop (PS). TPR is reported among brackets.}
	\vspace{0.1cm}
{\begin{tabular}{|p{1.5cm}<{\centering}|p{2cm}<{\centering}|p{2cm}<{\centering}|p{2cm}<{\centering}|}
            \hline
		   &$96\times96$ & $128\times128$&$156\times156$\\\hline
           \cite{bianchi2012image}-NAl & 0.285(0.35) &0.334(0.36)& 0.362(0.37)\\\hline
           Our & \textbf{0.534}(0.66)&\textbf{0.590}(0.71)&\textbf{0.620}(0.73)\\\hline
	\end{tabular}
	\label{tab.PS}
}
\end{table}

\begin{table}[t]
\renewcommand\arraystretch{1.2}
\centering
\scriptsize
\tiny

\caption{Localization performance (MCC) for the case  $128\times128$, when the first
compression is carried out with the PS software.}
 \begin{tabular}
{|c | c| c| c| c| c| c |c|}
  \hline
  \multicolumn{2}{|c|}{\diagbox[width=1.5cm]{$QF_{1}$}{$QF_{1,2}$}} &7&8& 9 & 10& 11&12 \\
  \hline  \hline
  \multirow{3}{*}{7}&\cite{bianchi2012image}-NAl&---&0.690(46)&0.686(39)&0.723(50)&0.698(54)&0.784(51)\\\cline{2-8}
  &Our&---&\textbf{0.721}(94)&\textbf{0.760}(97)&\textbf{0.784}(99)&\textbf{0.803}(97)&\textbf{0.802}(99)\\ \hline\hline
  \multirow{3}{*}{8}&\cite{bianchi2012image}-NAl&0.375(52)&---&0.364(52)&0.305(52)&0.356(50)&0.365(55)\\\cline{2-8}
  &Our&\textbf{0.484}(97)&---&\textbf{0.385}(72)&\textbf{0.489}(83)&\textbf{0.524}(87)&\textbf{0.566}(85)\\ \hline\hline
  \multirow{3}{*}{9}&\cite{bianchi2012image}-NAl&0.195(43)&0.150(37)&---&0.158(48)&0.168(42)&0.159(41)\\\cline{2-8}
  &Our&\textbf{0.654}(85)&\textbf{0.447}(33)&---&\textbf{0.208}(21)&\textbf{0.235}(21)&\textbf{0.198}(24)\\ \hline\hline
  \multirow{3}{*}{10}&\cite{bianchi2012image}-NAl&0.177(29)&0.157(28)&0.196(24)&---&\textbf{0.180}(28)&\textbf{0.168}(26)\\\cline{2-8}
  &Our&\textbf{0.755}(100)&\textbf{0.616}(96)&\textbf{0.354}(71)&---&0.115(32)&0.072(31)\\ \hline\hline
  \multirow{3}{*}{11}&\cite{bianchi2012image}-NAl&0.205(24)&0.186(26)&0.181(26)&0.194(29)&---&\textbf{0.198}(21)\\\cline{2-8}
  &Our&\textbf{0.786}(100)&\textbf{0.683}(98)&\textbf{0.532}(78)&\textbf{0.320}(44)&---&0.157(34)\\ \hline\hline
   \multirow{3}{*}{12}&\cite{bianchi2012image}-NAl&0.166(27)&0.229(25)&0.173(23)&0.183(25)&0.172(25)&---\\\cline{2-8}
  &Our&\textbf{0.802}(99)&\textbf{0.708}(99)&\textbf{0.595}(81)&\textbf{0.357}(53)&\textbf{0.211}(31)&---\\ \hline
  \end{tabular}%
  \label{tab:ps:128}

\end{table}

\begin{table*}[t]
\centering
\renewcommand\tabcolsep{3pt}
\scriptsize
\caption{Average localization performance of the methods (MCC) on the various datasets. TPR is reported among brackets.
}
	{\begin{tabular}{|c||c|c|c||c|c|c|c| c | c| c| c| c | c| c| c| c|}
            \hline
             & \multicolumn{3}{c||}{$\mathcal{D}_{ts}$}  & \multicolumn{3}{c|}{$\mathcal{D}_{Dr}$} &  \multirow{2}{*}{$\mathcal{D}_{{\scriptsize \cite{bianchi2012image}},\text{A}}$} &  \multirow{2}{*}{$\mathcal{D}_{{\scriptsize \cite{bianchi2012image}},\text{NA}}$} & \multirow{2}{*}{$\mathcal{D}_{{\scriptsize  \cite{wang2014exploring}}}$} &  \multicolumn{3}{c|}{$\mathcal{D}_{ts}'$} \\ 
             \cline{1-7} \cline{11-13}
               & All & Type I & Type II &  All &  Type I & { Type II} &  & & & All & Type I & Type II \\ \hline
            \cite{wu2019mantra}  & \textbf{0.651}(0.96) &\textbf{0.672} & \textbf{0.631} & 0.617(0.97) & 0.637& 0.595 & 0.411(0.37) & 0.119(0.33) & 0.328(0.72) &0.427(0.30) &0.474 &0.341\\ \hline
            Our  & 0.643(0.97) & 0.652&  0.633 & \textbf{0.631}(0.98)  & \textbf{0.639}& \textbf{0.622}  &  \textbf{0.904}(0.82) &\textbf{0.906}(0.80) &\textbf{0.712}(0.89) &  \textbf{0.624}(0.95)& \textbf{0.631}& \textbf{0.617} \\ \hline
	\end{tabular}}

	\label{tab.comparisonOnDBs}
\end{table*}

\subsection{Comparison with anomaly-based forgery detection}
\label{sec.MantraNet}


In this section, we discuss the performance achieved by the anomaly-based detector in \cite{wu2019mantra} (MantraNet), performing forgery localization by looking for general manipulation traces, and the comparison with our method.

The average localization performance achieved by MantraNet on the dataset $\mathcal{D}_{ts}$ and on the mismatched datastes are reported in Table \ref{tab.comparisonOnDBs}.
We observe that our method and MantraNet can achieve comparable performance on the dataset $\mathcal{D}_{ts}$, with the MantraNet method having a small advantage on Type I images (A-DJPEG case).

Similar results are achieved on the $\mathcal{D}_{Dr}$ dataset, with our method being slightly superior.

\begin{figure}[t]
\centering{\includegraphics[width =0.95\columnwidth]{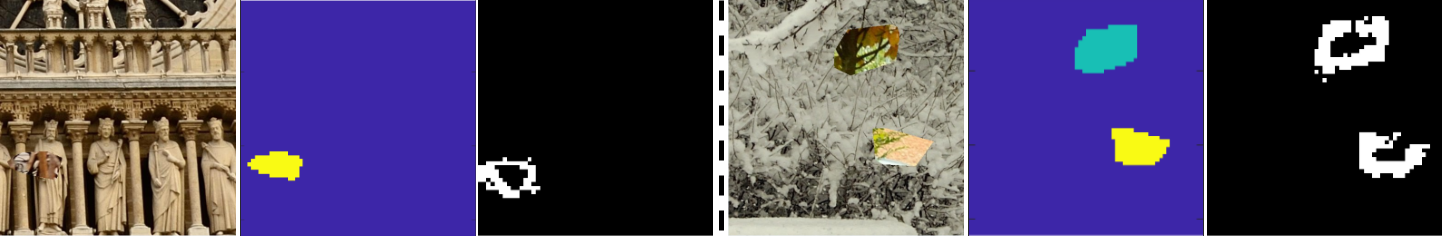}}
\caption{Examples of tampering maps from MantraNet \cite{wu2019mantra}.}
\label{fig.ExampleMantraNet}
\end{figure}

On the contrary, on the $\mathcal{D}_{{\scriptsize \cite{bianchi2012image}}}$ and $\mathcal{D}_{\scriptsize \cite{wang2014exploring}}$ dataset, the performance of MantraNet are significantly worse than those of our method.
Our explanation for this result is that MantraNet benefits from the way the tampered images are generated in our datasets.
In fact, the procedure of automatic generation of the tampered images in $\mathcal{D}_{ts}$ dataset, and  also in $\mathcal{D}_{Dr}$, leaves evident visual boundary artifacts clues in the images, with the shapes and the edges of the spliced regions being clearly visible, see the examples in Fig. \ref{fig:MRexamples}. This is
an obvious asset for a method like MantraNet that reveals the manipulation by looking at any kind of anomalies.  By inspecting  the localization maps we indeed see that MantraNet detects the border artifacts at the boundary of the spliced regions, see Fig. \ref{fig.ExampleMantraNet}.
On the datasets in  $\cite{bianchi2012image}$ and $\cite{wang2014exploring}$ (column 8-10 in Table \ref{tab.comparisonOnDBs}), where such visual artifacts are not present,  the performance of MantraNet are much worse, and also the detection performance are poor.
%
To further confirm this explanation, we also tested the various methods on a dataset $\mathcal{D}_{ts}'$ whose images are obtained as for $\mathcal{D}_{ts}$ with the only difference that, to avoid the boundary artifacts, for a given source image, the same images compressed with different quality factors are considered as donor images (similarly to what is done in $\cite{bianchi2012image}$ and $\cite{wang2014exploring}$).
The results we achieved (Table \ref{tab.comparisonOnDBs}, last three columns) confirms our expectation.
We can conclude that when the compression artifacts are
the only traces of tampering, an anomaly-based method looking for general
features like MantraNet might fail to detect and localize the tampering, thus confirming the advantage of resorting to dedicated solutions.

Last but not least, we stress that MantraNet can only localize the forgery, without distinguishing spliced regions from different donor images, that is, without performing attribution, which is a central goal of our method.

\section{Conclusion}
\label{sec.conclusion}

We have proposed an end-to-end system capable to detect and localize image splicing operations by additionally distinguishing spliced regions derived from different donor images (a task referred to as spliced regions attribution). The basic assumption behind the proposed system is that the spliced and background regions have been double JPEG-compressed, but the quantization matrices used for the first compression are different for the background and spliced regions stemming from different sources. Estimating the primary quantization matrix and clustering image blocks according to the result of the estimation provides the basic mechanism underlying the detection, localization and attribution of the spliced regions. We instantiated the proposed system by adopting state-of-the-art solutions for each step the system consists of, including estimation of the primary quantization matrix, estimation of the number of clusters, clustering and spatial refinement of the tampering map. The good performance of the resulting system have been proven by means of extensive experiments and compared with those of two baseline methods operating in similar conditions.
The main strength of the proposed system is that it can be applied to a wide variety of
situation, concerning the alignment of the double compression steps and the quantization steps used for the first and the second compression. Moreover, to the best of our knowledge, this is the first method that can distinguish multiple tampered regions taken from different donor images based on the analysis of DJPEG traces.

It goes without saying that any improvement of each of the steps the system consists of will result in a consequent improvement of the overall accuracy of the system. From this point of view, improving the accuracy of the estimation of the primary quantization matrix would play a crucial role, since the entire system relies on the accuracy of such a step. Clustering is also an area where improvements are possible, both with regard to the estimation of the number of clusters and the subsequent clustering process. In particular, finding better ways to fuse spatial information with the information provided by the estimated quantisation matrix could significantly improve the accuracy of the system.

\section*{Acknowledgments}

This work was supported by the National Natural Science Foundation of China (No. U1936212, 62120106009). This work has been partially supported by the PREMIER project under contract PRIN 2017 2017Z595XS-001, funded by the Italian Ministry of University and Research, and by the Defense Advanced Research Projects Agency (DARPA) and the Air Force Research Laboratory (AFRL) under agreement number FA8750-20-2-1004. The U.S. Government is authorized to reproduce and distribute reprints for Governmental purposes
notwithstanding any copyright notation thereon. The views and conclusions contained herein are those of the authors
and should not be interpreted as necessarily representing the official policies or endorsements, either expressed or implied,
of DARPA and AFRL or the U.S. Government.

\bibliographystyle{IEEEtran}
\bibliography{DJPEGQ1Loc}

\begin{thebibliography}{10}
\providecommand{\url}[1]{#1}
\csname url@samestyle\endcsname
\providecommand{\newblock}{\relax}
\providecommand{\bibinfo}[2]{#2}
\providecommand{\BIBentrySTDinterwordspacing}{\spaceskip=0pt\relax}
\providecommand{\BIBentryALTinterwordstretchfactor}{4}
\providecommand{\BIBentryALTinterwordspacing}{\spaceskip=\fontdimen2\font plus
\BIBentryALTinterwordstretchfactor\fontdimen3\font minus
  \fontdimen4\font\relax}
\providecommand{\BIBforeignlanguage}[2]{{%
\expandafter\ifx\csname l@#1\endcsname\relax
\typeout{** WARNING: IEEEtran.bst: No hyphenation pattern has been}%
\typeout{** loaded for the language `#1'. Using the pattern for}%
\typeout{** the default language instead.}%
\else
\language=\csname l@#1\endcsname
\fi
#2}}
\providecommand{\BIBdecl}{\relax}
\BIBdecl

\bibitem{Jessica2008}
T.~{Pevny} and J.~{Fridrich}, ``Detection of double-compression in {JPEG}
  images for applications in steganography,'' \emph{IEEE Trans. Inf. Forensics
  Security}, vol.~3, no.~2, pp. 247--258, June 2008.

\bibitem{Li2008}
{B. Li}, Y.~Q. {Shi}, and {J. Huang}, ``Detecting doubly compressed {JPEG}
  images by using mode based first digit features,'' in \emph{Proc. IEEE MMSP},
  Oct 2008, pp. 730--735.

\bibitem{amerini2014splicing}
I.~{Amerini}, R.~{Becarelli}, R.~{Caldelli}, and A.~D. {Mastio}, ``Splicing
  forgeries localization through the use of first digit features,'' in
  \emph{Proc. IEEE Int. Workshop Inf. Forensics Secur.}, 2014, pp. 143--148.

\bibitem{deng2019deep}
C.~{Deng}, Z.~{Li}, X.~{Gao}, and D.~{Tao}, ``Deep multi-scale discriminative
  networks for double {JPEG} compression forensics,'' \emph{{ACM} Trans.
  Intell. Syst. Technol.}, vol.~10, no.~2, pp. 1--20, 2019.

\bibitem{chen2011detecting}
Y.-L. {Chen} and C.-T. {Hsu}, ``Detecting recompression of {JPEG} images via
  periodicity analysis of compression artifacts for tampering detection,''
  \emph{IEEE Trans. Inf. Forensics Security}, vol.~6, no.~2, pp. 396--406,
  2011.

\bibitem{niu2019SPL}
Y.~{Niu}, B.~{Tondi}, Y.~{Zhao}, and M.~{Barni}, ``Primary quantization matrix
  estimation of double compressed {JPEG} images via {CNN},'' \emph{IEEE Signal
  Process. Lett.}, vol.~27, pp. 191--195, 2020.

\bibitem{von2007tutorial}
U.~Von~Luxburg, ``A tutorial on spectral clustering,'' \emph{Statistics and
  computing}, vol.~17, no.~4, pp. 395--416, 2007.

\bibitem{mahdian2008blind}
B.~{Mahdian} and S.~{Saic}, ``Blind authentication using periodic properties of
  interpolation,'' \emph{IEEE Trans. Inf. Forensics Security}, vol.~3, no.~3,
  pp. 529--538, 2008.

\bibitem{bunk2017detection}
J.~{Bunk}, J.~H. {Bappy}, T.~M. {Mohammed}, L.~{Nataraj}, A.~{Flenner},
  B.~{Manjunath}, S.~{Chandrasekaran}, A.~K. {Roy-Chowdhury}, and
  L.~{Peterson}, ``Detection and localization of image forgeries using
  resampling features and deep learning,'' in \emph{Proc. IEEE Conf. Comput.
  Vis. Pattern Recognit Workshops (CVPRW)}, 2017, pp. 1881--1889.

\bibitem{chierchia2014guided}
G.~{Chierchia}, D.~{Cozzolino}, G.~{Poggi}, C.~{Sansone}, and L.~{Verdoliva},
  ``Guided filtering for {PRNU}-based localization of small-size image
  forgeries,'' in \emph{Proc. IEEE Int. Conf. Acoust., Speech Signal Process.},
  2014, pp. 6231--6235.

\bibitem{korus2017multi}
P.~{Korus} and J.~{Huang}, ``Multi-scale analysis strategies in {PRNU}-based
  tampering localization,'' \emph{IEEE Trans. Inf. Forensics Security},
  vol.~12, no.~4, pp. 809--824, 2017.

\bibitem{popescu2004statistical}
A.~C. Popescu and H.~Farid, ``Statistical tools for digital forensics,'' in
  \emph{Proc. 6th Int. workshop on Inf. Hiding}, 2004, pp. 128--147.

\bibitem{pasquini2014multiple}
C.~Pasquini, G.~Boato, and F.~P{\'e}rez-Gonz{\'a}lez, ``Multiple {JPEG}
  compression detection by means of {B}enford-{F}ourier coefficients,'' in
  \emph{Proc. IEEE Int. Workshop Inf. Forensics Secur.}, 2014, pp. 113--118.

\bibitem{luo2007BACM}
W.~{Luo}, Z.~{Qu}, J.~{Huang}, and G.~{Qiu}, ``A novel method for detecting
  cropped and recompressed image block,'' in \emph{Proc. IEEE Int. Conf.
  Acoust., Speech Signal Process.}, vol.~2, 2007, pp. 217--220.

\bibitem{qu2008a}
Z.~{Qu}, W.~{Luo}, and J.~{Huang}, ``A convolutive mixing model for shifted
  double {JPEG} compression with application to passive image authentication,''
  in \emph{Proc. IEEE Int. Conf. Acoust., Speech Signal Process.}, 2008, pp.
  1661--1664.

\bibitem{bianchi2012detection}
T.~{Bianchi} and A.~{Piva}, ``Detection of nonaligned double {JPEG} compression
  based on integer periodicity maps,'' \emph{IEEE Trans. Inf. Forensics
  Security}, vol.~7, no.~2, pp. 842--848, 2012.

\bibitem{lin2009fast}
Z.~{Lin}, J.~{He}, X.~{Tang}, and C.-K. {Tang}, ``Fast, automatic and
  fine-grained tampered {JPEG} image detection via {DCT} coefficient
  analysis,'' \emph{Pattern Recognition}, vol.~42, no.~11, pp. 2492--2501,
  2009.

\bibitem{bianchi2011improved}
T.~{Bianchi}, A.~D. {Rosa}, and A.~{Piva}, ``Improved dct coefficient analysis
  for forgery localization in {JPEG} images,'' in \emph{Proc. IEEE Int. Conf.
  Acoust., Speech Signal Process.}, 2011, pp. 2444--2447.

\bibitem{WZ16}
Q.~Wang and R.~Zhang, ``Double {JPEG} compression forensics based on a
  convolutional neural network,'' \emph{{EURASIP} Journal on Information
  Security}, vol. 2016, no.~1, pp. 1--12, 2016.

\bibitem{barni2017aligned}
M.~Barni, L.~Bondi, N.~Bonettini, P.~Bestagini, A.~Costanzo, M.~Maggini,
  B.~Tondi, and S.~Tubaro, ``Aligned and non-aligned double {JPEG} detection
  using convolutional neural networks,'' \emph{Journal of Visual Communication
  and Image Representation}, vol.~49, pp. 153--163, 2017.

\bibitem{Kwon2021WACV}
M.-J. Kwon, I.-J. Yu, S.-H. Nam, and H.-K. Lee, ``Cat-net: Compression artifact
  tracing network for detection and localization of image splicing,'' in
  \emph{Proceedings of the IEEE/CVF Winter Conference on Applications of
  Computer Vision}, 2021, pp. 375--384.

\bibitem{Lukas2003}
J.~Luk\'{a}\v{s} and J.~Fridrich, ``Estimation of primary quantization matrix
  in double compressed {JPEG} images,'' \emph{Proc. Digital Forensic Research
  Workshop}, 2003.

\bibitem{Farid2009}
H.~{Farid}, ``Exposing digital forgeries from {JPEG} ghosts,'' \emph{IEEE
  Trans. Inf. Forensics Security}, vol.~4, no.~1, pp. 154--160, March 2009.

\bibitem{bianchi2012image}
T.~Bianchi and A.~Piva, ``Image forgery localization via block-grained analysis
  of jpeg artifacts,'' \emph{IEEE Trans. Inf. Forensics Security}, vol.~7,
  no.~3, pp. 1003--1017, 2012.

\bibitem{wang2014exploring}
W.~Wang, J.~Dong, and T.~Tan, ``Exploring {DCT} coefficient quantization
  effects for local tampering detection,'' \emph{IEEE Trans. Inf. Forensics
  Security}, vol.~9, no.~10, pp. 1653--1666, 2014.

\bibitem{ImagePhylogeny}
Z.~{Dias}, A.~{Rocha}, and S.~{Goldenstein}, ``First steps toward image
  phylogeny,'' in \emph{Proc. IEEE Int. Workshop Inf. Forensics Secur.}, 2010,
  pp. 1--6.

\bibitem{RochaTIFS2011}
------, ``Image phylogeny by minimal spanning trees,'' \emph{IEEE Trans. Inf.
  Forensics Security}, vol.~7, no.~2, pp. 774--788, 2012.

\bibitem{Oliveira2016ImagePhylogeny}
A.~A. {de Oliveira}, P.~{Ferrara}, A.~{De Rosa}, A.~{Piva}, M.~{Barni},
  S.~{Goldenstein}, Z.~{Dias}, and A.~{Rocha}, ``Multiple parenting phylogeny
  relationships in digital images,'' \emph{IEEE Trans. Inf. Forensics
  Security}, vol.~11, no.~2, pp. 328--343, 2016.

\bibitem{wu2018busternet}
Y.~Wu, W.~Abd-Almageed, and P.~Natarajan, ``Busternet: Detecting copy-move
  image forgery with source/target localization,'' in \emph{Proceedings of the
  European Conference on Computer Vision (ECCV)}, 2018, pp. 168--184.

\bibitem{Tin}
M.~Barni, Q.-T. Phan, and B.~Tondi, ``Copy move source-target disambiguation
  through multi-branch cnns,'' \emph{IEEE Transactions on Information Forensics
  and Security}, vol.~16, pp. 1825--1840, 2021.

\bibitem{islam2020doa}
A.~Islam, C.~Long, A.~Basharat, and A.~Hoogs, ``Doa-gan: Dual-order attentive
  generative adversarial network for image copy-move forgery detection and
  localization,'' in \emph{Proceedings of the IEEE/CVF Conference on Computer
  Vision and Pattern Recognition}, 2020, pp. 4676--4685.

\bibitem{zhou2018learning}
P.~Zhou, X.~Han, V.~I. Morariu, and L.~S. Davis, ``Learning rich features for
  image manipulation detection,'' in \emph{Proceedings of the IEEE Conference
  on Computer Vision and Pattern Recognition}, 2018, pp. 1053--1061.

\bibitem{wu2019mantra}
Y.~Wu, W.~AbdAlmageed, and P.~Natarajan, ``Mantra-net: Manipulation tracing
  network for detection and localization of image forgeries with anomalous
  features,'' in \emph{Proceedings of the IEEE/CVF Conference on Computer
  Vision and Pattern Recognition}, 2019, pp. 9543--9552.

\bibitem{marra2020full}
F.~Marra, D.~Gragnaniello, L.~Verdoliva, and G.~Poggi, ``A full-image
  full-resolution end-to-end-trainable cnn framework for image forgery
  detection,'' \emph{IEEE Access}, vol.~8, pp. 133\,488--133\,502, 2020.

\bibitem{pennebaker1992jpeg}
W.~B. Pennebaker and J.~L. Mitchell, \emph{JPEG: Still image data compression
  standard}.\hskip 1em plus 0.5em minus 0.4em\relax Springer Science \&
  Business Media, 1992.

\bibitem{amerini2014blind}
I.~Amerini, R.~Caldelli, P.~Crescenzi, A.~Del~Mastio, and A.~Marino, ``Blind
  image clustering based on the normalized cuts criterion for camera
  identification,'' \emph{Signal Process., Image Commun.}, vol.~29, no.~8, pp.
  831--843, 2014.

\bibitem{li2017mobile}
Y.~Li, X.~Zhang, X.~Li, Y.~Zhang, J.~Yang, and Q.~He, ``Mobile phone clustering
  from speech recordings using deep representation and spectral clustering,''
  \emph{IEEE Trans. Inf. Forensics Security}, vol.~13, no.~4, pp. 965--977,
  2017.

\bibitem{mayer2019exposing}
O.~Mayer and M.~C. Stamm, ``Exposing fake images with forensic similarity
  graphs,'' \emph{IEEE J. Sel. Topics Signal Process.}, vol.~14, no.~5, pp.
  1049--1064, 2020.

\bibitem{mclachlan1988mixture}
G.~J. {McLachlan} and K.~E. {Basford}, \emph{Mixture models: inference and
  applications to clustering}, 1988, vol.~84.

\bibitem{Vesanto2000}
J.~{Vesanto} and E.~{Alhoniemi}, ``Clustering of the self-organizing map,''
  \emph{IEEE Trans. Neural Netw}, vol.~11, no.~3, pp. 586--600, 2000.

\bibitem{1998fuzzy}
Y.~{El-Sonbaty} and M.~{Ismail}, ``Fuzzy clustering for symbolic data,''
  \emph{IEEE Trans. Fuzzy Syst.}, vol.~6, no.~2, pp. 195--204, 1998.

\bibitem{simonyan2014very}
K.~Simonyan and A.~Zisserman, ``Very deep convolutional networks for
  large-scale image recognition,'' \emph{arXiv preprint arXiv:1409.1556}, 2014.

\bibitem{gonzales2002digital}
R.~C. Gonzales and R.~E. Woods, ``Digital image processing,'' 2002.

\bibitem{matthews1975}
B.~W. {Matthews}, ``Comparison of the predicted and observed secondary
  structure of t4 phage lysozyme.'' \emph{Biochimica et Biophysica Acta}, vol.
  405, no.~2, pp. 442--451, 1975.

\bibitem{dataMining}
P.-N. Tan, M.~Steinbach, and V.~Kumar, \emph{Introduction to Data
  Mining}.\hskip 1em plus 0.5em minus 0.4em\relax USA: Addison-Wesley Longman
  Publishing Co., Inc., 2005.

\bibitem{RAISE8K}
\BIBentryALTinterwordspacing
D.~Dang-Nguyen, C.~Pasquini, V.~Conotter, and G.~Boato, ``{RAISE}: A raw images
  dataset for digital image forensics,'' in \emph{Proc. the 6th ACM Multimedia
  Systems Conference}, 2015, pp. 219--224. [Online]. Available:
  \url{http://doi.acm.org/10.1145/2713168.2713194}
\BIBentrySTDinterwordspacing

\bibitem{UCID}
G.~Schaefer and M.~Stich, ``{UCID}: An uncompressed color image database,'' in
  \emph{Storage and Retrieval Methods and Applications for Multimedia 2004},
  vol. 5307, 2003, pp. 472--480.

\bibitem{Dresden}
\BIBentryALTinterwordspacing
T.~Gloe and R.~B\"{o}hme, ``The '{D}resden {I}mage {D}atabase' for benchmarking
  digital image forensics,'' in \emph{Proceedings of the 2010 ACM Symposium on
  Applied Computing}, ser. SAC '10.\hskip 1em plus 0.5em minus 0.4em\relax New
  York, NY, USA: ACM, 2010, pp. 1584--1590. [Online]. Available:
  \url{http://doi.acm.org/10.1145/1774088.1774427.}
\BIBentrySTDinterwordspacing

\end{thebibliography}

\end{document}